\documentclass{emulateapj}
\usepackage{natbib,aas_macros}
\usepackage{multirow,color}

\begin{document}
\shorttitle{A Close Comparison between Observed and Modeled Ly$\alpha$ Lines}
\shortauthors{Hashimoto et al.}
\slugcomment{Ver. Aug. 24, 2015}

\title{
A Close Comparison between Observed and Modeled Ly$\alpha$ Lines \\
for $z\sim2.2$ Lyman Alpha Emitters
\altaffilmark{\dag}
\altaffilmark{\ddag}
\altaffilmark{\dag \dag}
}
\author{
Takuya Hashimoto ~\altaffilmark{1},
Anne Verhamme~\altaffilmark{2},
Masami Ouchi~\altaffilmark{3,4},
Kazuhiro Shimasaku~\altaffilmark{1,5}, 
Daniel Schaerer~\altaffilmark{2,6},
Kimihiko Nakajima~\altaffilmark{2},
Takatoshi Shibuya~\altaffilmark{3},
Michael Rauch~\altaffilmark{7},
Yoshiaki Ono~\altaffilmark{3,4}, 
and 
Ryosuke Goto~\altaffilmark{1}.
}

\email{thashimoto \_at\_ astron.s.u-tokyo.ac.jp}

\altaffiltext{1}{Department of Astronomy, Graduate School of Science, The University of Tokyo, Tokyo 113-0033, Japan}
\altaffiltext{2}{Observatoire de Gen\`eve, Universit\'e de Gen\`eve, 51 Ch. des Maillettes, 1290 Versoix, Switzerland}
\altaffiltext{3}{Institute for Cosmic Ray Research, The University of Tokyo, 5-1-5 Kashiwanoha, Kashiwa, Chiba 277-8582, Japan}
\altaffiltext{4}{Kavli Institute for the Physics and Mathematics of the Universe (WPI),The University of Tokyo, 5-1-5 Kashiwanoha, Kashiwa, Chiba 277-8583, Japan}
\altaffiltext{5}{Research Center for the Early Universe, Graduate School of Science, The University of Tokyo, Tokyo 113-0033, Japan}
\altaffiltext{6}{CNRS, IRAP, 14 Avenue E. Belin, 31400 Toulouse, France}
\altaffiltext{7}{Observatories of the Carnegie Institution of Washington, 813 Santa Barbara Street, Pasadena, CA 91101, USA}

\altaffiltext{\dag}{
Some of the data presented herein were obtained at the W. M. Keck 
Observatory, which is operated as a scientific partnership 
among the California Institute of Technology, 
the University of California, 
and the National Aeronautics and Space Administration. 
The Observatory was made possible by the generous financial 
support of the W. M. Keck Foundation.
}

\altaffiltext{\ddag}{
Based in part on data collected at the Subaru Telescope,
which is operated by the National Astronomical Observatory of Japan.
}

\altaffiltext{\dag\dag}{
This paper includes data gathered with the 6.5 meter Magellan Telescopes
located at Las Campanas Observatory, Chile.
}

\slugcomment{}

\begin{abstract}
We present the results of a Ly$\alpha$ profile analysis of 12 
Ly$\alpha$ emitters (LAEs) at $z \sim 2.2$ with high-resolution
Ly$\alpha$ spectra.
We find that all 12 objects have a Ly$\alpha$ profile with the main
peak redward of the systemic redshift defined by nebular lines, 
and five have a weak, secondary peak blueward of the systemic
redshift (blue bump).
The average velocity offset of the red main peak (the blue bump, 
if any) with respect to the systemic redshift is 
$\Delta v_{\rm Ly\alpha,r} = 174 \pm 19$ km s$^{-1}$ 
($\Delta v_{\rm Ly\alpha,b} = -316 \pm 45$ km s$^{-1}$), 
which is smaller than (comparable to) that of Lyman-break galaxies (LBGs).
The outflow velocities inferred from metal absorption lines in 
three individual and one stacked spectra are comparable to those 
of LBGs.
The uniform expanding shell model constructed by \cite{verhamme2006} 
reproduces not only the Ly$\alpha$ profiles but also other observed 
quantities including the outflow velocity and the FWHM of nebular
lines for the non-blue bump objects.
On the other hand, the model predicts too high FWHMs of nebular
lines for the blue bump objects, although this 
discrepancy may disappear if we introduce additional Ly$\alpha$ 
photons produced by gravitational cooling.
We show that the small $\Delta v_{\rm Ly\alpha,r}$ values of our sample can be 
explained by low neutral-hydrogen column densities of
log($N_{\rm HI}$) = 18.9 cm$^{-2}$ on average.
This value is more than one order of magnitude lower 
than those of LBGs but is consistent with recent findings that 
LAEs have high ionization parameters and low {\sc Hi} gas masses. 
This result suggests that low  $N_{\rm HI}$ values, giving reduced 
numbers of resonant scattering of Ly$\alpha$ photons, are 
the key to the strong Ly$\alpha$ emission of LAEs.
\end{abstract}

\keywords{
galaxies: high-redshift ---
galaxies: ISM ---
line: profiles --- radiative transfer 
}

\section{INTRODUCTION} \label{sec:introduction}

Ly$\alpha$ emitters (LAEs) are objects commonly seen in both the local and high-$z$ universes 
with large Ly$\alpha$ equivalent widths, EW(Ly$\alpha$)$\gtrsim20-30$ \AA 
(local: \citealt{deharveng2008, cowie2011},
high-$z$: \citealt{hu1996, rhoads2001, ouchi2008, ouchi2010}). 
Previous studies based on Spectral Energy Distributions (SEDs) have revealed that 
typical LAEs are young, low-mass galaxies with a small dust content 
(e.g., \citealt{nilsson2007, gawiser2007, guaita2011, nakajima2012, kusakabe2015}), 
although there are some evolved LAEs with a moderate mass and dust (\citealt{ono2010a, hagen2014}). 
Morphological studies of their UV continuum have shown that 
the galactic counterparts of LAEs to be typically compact (e.g., \citealt{bond2009})
and their typical size does not evolve with redshift (\citealt{malhotra2012}). 
Furthermore, clustering analyses have revealed that LAEs have the 
lowest dark matter halo masses at every redshift (\citealt{ouchi2010, guaita2010}). 
These properties suggest that LAE is an important galaxy population as the 
building block candidates in the $\Lambda$ CDM model (\citealt{rauch2008}).

Given their importance in galaxy evolution, 
the Ly$\alpha$ escape mechanism in LAEs is still poorly understood.
Resonant scattering strongly extends the path-length of Ly$\alpha$ photons 
through galactic gas 
and renders them prone to absorption by dust grains. 
On one hand, 
some observational studies at the local universe have proposed that 
outflows facilitate the escape of Ly$\alpha$ photons from galaxies (e.g., \citealt{kunth1998})
as they reduce the number of scattering. 
Likewise, others  (e.g., \citealt{kornei2010, atek2014}) have 
shown that the dust content correlates with Ly$\alpha$ emissivity. 
While these effects would certainly be at work, there has been no decisive conclusion (cf., \citealt{cassata2015}). 
On the other hand, 
theoretical studies have computed the Ly$\alpha$ radiation transfer (RT) 
through idealized spherically symmetric shells of homogeneous and 
isothermal neutral hydrogen gas, especially in a form of an expanding shell  
(e.g., \citealt{zheng2002, verhamme2006, dijkstra2009, kollmeier2010}). 
They have investigated 
how properties of the interstellar medium (ISM) affect the Ly$\alpha$ escape and 
emergent Ly$\alpha$ profiles. 
The result is that the Ly$\alpha$ RT 
is a complicated process altered by 
galactic outflows/inflows, 
the neutral hydrogen column density and dust content of the ISM, 
and the inclination of the galaxy disk (e.g., \citealt{verhamme2012, zheng2014, behrens2014b}). 
One of the goals in these theoretical studies is 
to aid understanding the galaxy properties from observed Ly$\alpha$ lines, 
and to identify the key factor for the Ly$\alpha$ escape.

To study the Ly$\alpha$ RT and escape through close comparisons of 
observed and modeled Ly$\alpha$ lines, 
it is important to obtain spectral lines other than the Ly$\alpha$ line. 
The central wavelength and the width of nebular lines 
(e.g., H$\alpha$ and [{\sc Oiii}])
tell us the galaxy's systemic redshift 
and internal velocity. 
The blue-shift of interstellar (IS) absorption lines 
with respect to the systemic redshift gives the galactic outflow velocity, 
and the width of the IS lines can be interpreted as 
the sum of thermal and macroscopic (rotation and turbulence) velocities of the outflowing gas. 
These lines can help us to disentangle the
complicated Ly$\alpha$ RT and understand the Ly$\alpha$ escape.

However, due to the typical faintness of LAEs, 
it is only recently that these additional lines have been successfully detected 
in narrow-band selected LAEs 
(nebular lines: e.g., \citealt{mclinden2011, finkelstein2011b, hashimoto2013}, 
IS absorption lines: \citealt{hashimoto2013, shibuya2014b}). 
Thus, in contrast to LBGs whose Ly$\alpha$ profiles have been closely compared with 
Ly$\alpha$ RT models (e.g., \citealt{verhamme2008, kulas2012, christensen2012}), 
there are only a few studies that have performed 
Ly$\alpha$ profile comparisons of LAEs (e.g., \citealt{chonis2013}). 
Recent simultaneous detections of Ly$\alpha$ and nebular emission lines 
have statistically confirmed that 
the Ly$\alpha$ profiles of LAEs are asymmetric with a red main peak 
redshifted with respect to the systemic 
$\Delta v_{\rm Ly\alpha,r}$, $>$  0 km s$^{-1}$ 
(e.g., \citealt{shibuya2014b, song2014, erb2014}). 
Likewise, IS absorption studies in LAEs have shown that they are blue-shifted 
with respect to the systemic by $|\Delta v_{\rm abs}| \sim 100 - 200$ km s$^{-1}$ (\citealt{hashimoto2013, shibuya2014b}), 
which is comparable to those of LBGs (e.g., \citealt{pettini2001, shapley2003, steidel2010, kulas2012}). 
These results suggest that LAEs do have outflows 
and motivate us to apply expanding shell models to LAEs.

To examine Ly$\alpha$ escape mechanisms in LAEs through detailed Ly$\alpha$ modeling, 
we focus on the small  $\Delta v_{\rm Ly\alpha,r}$ of LAEs, 
$\Delta v_{\rm Ly\alpha,r} \simeq $ 200 km s$^{-1}$, 
 compared to those of LBGs,  
$\Delta v_{\rm Ly\alpha,r} \simeq$ 400 km s$^{-1}$ (e.g., \citealt{steidel2010, kulas2012}),  
with similar physical quantities such as stellar mass, star formation rate (SFR), 
and velocity dispersion 
(\citealt{hashimoto2013, shibuya2014b, song2014, erb2014}). 
\cite{hashimoto2013} and \cite{shibuya2014b} have also shown that 
LAEs have comparable outflow velocities, measured from IS absorption lines, 
to those of LBGs. 
These results imply that a definitive difference between LAEs and LBGs 
in velocity properties is $\Delta v_{\rm Ly\alpha,r}$. 
In addition, \cite{hashimoto2013} have demonstrated that 
EW(Ly$\alpha$) anti-correlates with $\Delta v_{\rm Ly\alpha,r}$ 
using a large sample of LAEs and LBGs (see also \citealt{shibuya2014b, erb2014}). 
Therefore, 
understanding the reason why LAEs have small $\Delta v_{\rm Ly\alpha,r}$,  
through  detailed Ly$\alpha$ modeling 
should shed light on the Ly$\alpha$ RT and Ly$\alpha$ escape mechanisms in LAEs.

According to the theoretical studies, there are several possible explanations  
for a small $\Delta v_{\rm Ly\alpha,r}$: 
a high outflow velocity (e.g., \citealt{verhamme2006}), 
a very low neutral hydrogen column density ($N_{\rm HI}$) of the ISM 
(e.g., \citealt{verhamme2006, verhamme2015}), 
an inhomogeneous ISM with a covering fraction ($CF$) below unity, 
where $CF$ is defined as the fraction of sightlines  
which are optically thick to the Ly$\alpha$ radiation, 
i.e., gas with holes (e.g., \citealt{behrens2014a, verhamme2015}), 
and a clumpy ISM with a low covering factor, $f_{\rm c}$, 
which is defined as the average number of clouds intersected by a random line of sight 
(e.g., \citealt{hansen_oh2006, dijkstra2012, laursen2013})

In this work, we focus on applying the uniform expanding shell model based on a 3D Ly$\alpha$ RT 
constructed by \cite{verhamme2006} and \cite{schaerer2011}, 
to 12 LAEs 
whose Ly$\alpha$ and nebular emission lines (e.g., H$\alpha$, {\sc Oiii}) 
have been observed at a high spectral resolution (\citealt{hashimoto2013, nakajima2013, shibuya2014b}). 
With the systemic redshifts and the full width half maximums (FWHM) 
determined from nebular emission lines, 
the stellar dust extinction derived from SED fitting, 
and the galactic outflow velocities inferred from LIS absorption lines,  
we first statistically examine how well the model can 
reproduce the Ly$\alpha$ profiles and other observables 
(cf., \citealt{verhamme2008, kulas2012, chonis2013}). 
After demonstrating the validity of the model, 
we securely derive physical quantities such as $N_{\rm HI}$ 
and discuss the origin of the small $\Delta v_{\rm Ly\alpha,r}$ 
and implications for the Ly$\alpha$ escape in LAEs. 
Possible other scenarios mentioned above for the small $\Delta v_{\rm Ly\alpha,r}$ 
are also qualitatively discussed.

This paper is organized as follows. 
We describe our spectroscopy observations in Section \ref{sec:data_obs}, 
and discuss profiles of Ly$\alpha$ and nebular emission lines 
in Section \ref{sec:obs_results}. 
We apply the uniform expanding shell model to our data 
and show comparisons with observables in Section \ref{sec:model_fitting}.
Discussion on the blue bumps as well as the origin of the small Ly$\alpha$ velocity offsets 
are given in Section \ref{sec:discussion}, 
followed by conclusions in Section \ref{sec:conclusions}.

Throughout this paper, magnitudes are given in the AB system
\citep{oke1983}, and we assume a $\Lambda$CDM cosmology 
with $\Omega_{\small m} = 0.3$, $\Omega_{\small \Lambda} = 0.7$
and $H_{\small 0} = 70$ km s$^{-1}$ Mpc$^{-1}$.

\section{Data and Observations} \label{sec:data_obs}

Our initial sample of objects are taken from large $z\sim2.2$ LAE samples 
in the COSMOS field, the Chandra Deep Field South (CDFS), 
and the Subaru/{\it XMM-Newton} Deep Survey (SXDS) 
(\citealt{nakajima2012, nakajima2013}; Nakajima et al. in prep.). 
These LAE samples are all based on 
Subaru/Suprime-Cam imaging observations 
with our custom made narrow band filter, 
NB387 ($\lambda_{\rm c}$ = 3870\AA\ and FWHM = 94\AA). 
The LAEs have been selected by color criteria of 
$B$ $-$ NB387 and $u^{*}$ $-$ NB387, 
satisfying the condition that the 
rest frame photometric Ly$\alpha$ EW (EW(Ly$\alpha$)$_{\rm photo}$)
be larger than 30\AA. 
From these, we only use 12 LAEs 
whose Ly$\alpha$ and nebular emission lines (e.g., H$\alpha$ and [O{\sc iii}]) 
are both spectroscopically confirmed. 
Among the  12 objects, 11 have been presented in \cite{hashimoto2013} and \cite{shibuya2014b}. 
We add one new LAE with EW(Ly$\alpha$)$_{\rm photo}$ $\sim$ 280\AA\, 
whose detailed properties will be discussed in Hashimoto et al. in prep.

In this section, 
we briefly summarize 
our near-infrared spectroscopy (\S \ref{subsec:nir_observation}),  
optical spectroscopy (\S \ref{subsec:optical_observation}), 
and the contamination of AGNs in the sample (\S \ref{subsec:agn}).

\subsection{Near-Infrared Specctroscopy} \label{subsec:nir_observation}

In order to detect nebular emission lines, 
we performed three near-infrared observations 
with Magellan/MMIRS (PI: M. Ouchi), Keck/NIRSPEC (PI: K. Nakajima), and Subaru/FMOS (PI: K. Nakajima). 
Canonical spectral resolutions for our observation settings
are $R\sim$ 1120, 1500, and 2200 for MMIRS, NIRSPEC, and FMOS, respectively. 

Details of the observation and data reduction procedures for MMIRS and NIRSPEC 
have been presented in \cite{hashimoto2013} and \cite{nakajima2013}, respectively. 
Briefly, two CDFS objects, CDFS-3865 and CDFS-6482, were observed with MMIRS 
using the HK grism covering $2.254-2.45$ $\mu$m, 
resulting in successful H$\alpha$ and [{\sc Oiii}]$\lambda \lambda$ $4959$ $5007$ detections. 
A follow-up observation was carried out for CDFS-3865 with NIRSPEC. 
The [{\sc Oii}] $\lambda 3727$ line was additionally detected 
with the $J$ band ($1.15-1.36$ $\mu$m). 
Four COSMOS objects, 
COSMOS-08501, COSMOS-13636, COSMOS-30679, and COSMOS-43982, 
were observed with NIRSPEC and its $K$ band ($2.2-2.43$ $\mu$m),  
resulting in H$\alpha$ line detections. 
The [{\sc Oiii}] $\lambda5007$ line was also detected from COSMOS-30679 
using the $H$ band ($1.48-1.76$ $\mu$m). 

The data from  FMOS will be presented in Nakajima et al. (2015, in prep). 
Its spectral coverage is $0.9-1.8$ $\mu$m. 
We detected [{\sc Oiii}] line(s) in eight objects: COSMOS-08357, COSMOS-12805, 
COSMOS-13138, COSMOS-13636, COSMOS-38380, COSMOS-43982, SXDS-10600, and SXDS-10942. 

\subsection{Optical Spectroscopy} \label{subsec:optical_observation}

In order to detect Ly$\alpha$ and metal absorption lines, 
we carried out several observations 
with Magellan/MagE (PI: M. Rauch) and Keck/LRIS (PI: M. Ouchi). 
The spectral resolutions for our observations 
were $R\sim4100$ and $\sim1100$ for MagE and LRIS, respectively. 
The slit was positioned on the Ly$\alpha$ centroids in the NB387 images.

Details of the observation and data reduction procedures for MagE and LRIS 
have been presented in \cite{hashimoto2013} and \cite{shibuya2014b}, respectively, 
except for COSMOS-08501. 
First, we describe this new object in detail (\S \ref{subsec:optical_observation_c08501})
and then giver a brief summary for the rest of the sample (\S \ref{subsec:optical_observation_rest}). 

\subsubsection{Optical Spectroscopy for COSMOS-08501} \label{subsec:optical_observation_c08501}

The MagE observations were carried out for COSMOS-08501 on 2012 February and 2013 December.  
We obtained $3\times3000$ s and $1\times3000$ s exposure times during each run, 
resulting in a 12000 s total integration time. 
Spectroscopic standard stars, dome flats, and Xenon flash lamp flats, 
were obtained on each night for calibrations. 
On these nights, the typical seeing sizes were $1''.0$.
The slit width was $1.''0$ for both runs, corresponding to $R\sim4100$. 
The spectra were reduced with $\tt{IDL}$ based pipeline, $\tt{MagE\_REDUCE}$, 
constructed by G. Becker (see also \citealt{kelson2003}). 
This pipeline processes raw frames, 
performing wavelength calibration and optimal sky subtraction,  
and extracts 1D spectra. 
Each of these reduced frames was then combined to form our final calibrated spectrum. 
From this, the Ly$\alpha$ line was identified above the 3 $\sigma$ noise 
of the local continuum. 

\subsubsection{Optical Spectroscopy for the Rest of the Sample} \label{subsec:optical_observation_rest}

CDFS-3865, CDFS-6482, and COSMOS-30679 were observed with MagE; 
COSMOS-08357, COSMOS-12085, COSMOS-13138, COSMOS-38380, SXDS-10600, and SXDS-10942 
were observed with LRIS; and finally, COSMOS-13636 and COSMOS-43982 were observed 
with both spectrographs. 
We identified the Ly$\alpha$ line in all objects. 
In addition, we detected several metal absorption lines 
(e.g., {\sc Si ii} $\lambda 1260$ and {\sc Civ} $\lambda1548$ lines) 
in a stacked MagE spectrum of CDFS-3865, CDFS-6482, COSMOS-13636, 
and COSMOS-30679 (\citealt{hashimoto2013})
as well as in individual LRIS spectra of 
COSMOS-12805, COSMOS-13636, and SXDS-10600 (\citealt{shibuya2014b}). 

A summary of our observations is listed in Table  \ref{tab:summary_observations}, 
and our Ly$\alpha$ and nebular emission line profiles are shown in Figure 1.

\begin{figure*}[]
\centering
\label{fig:testtest}
\includegraphics[width=19cm]{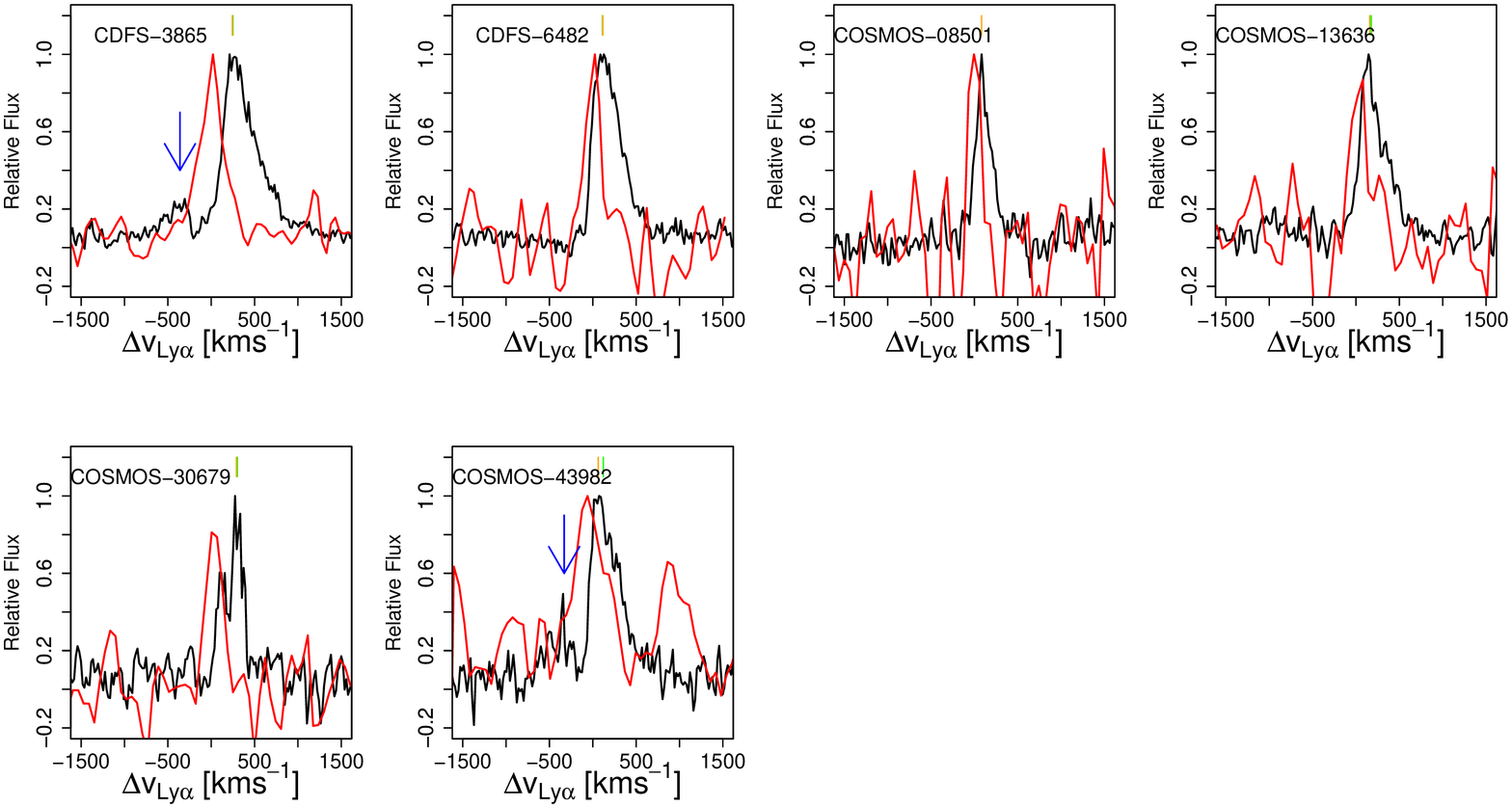} 
\includegraphics[width=19cm]{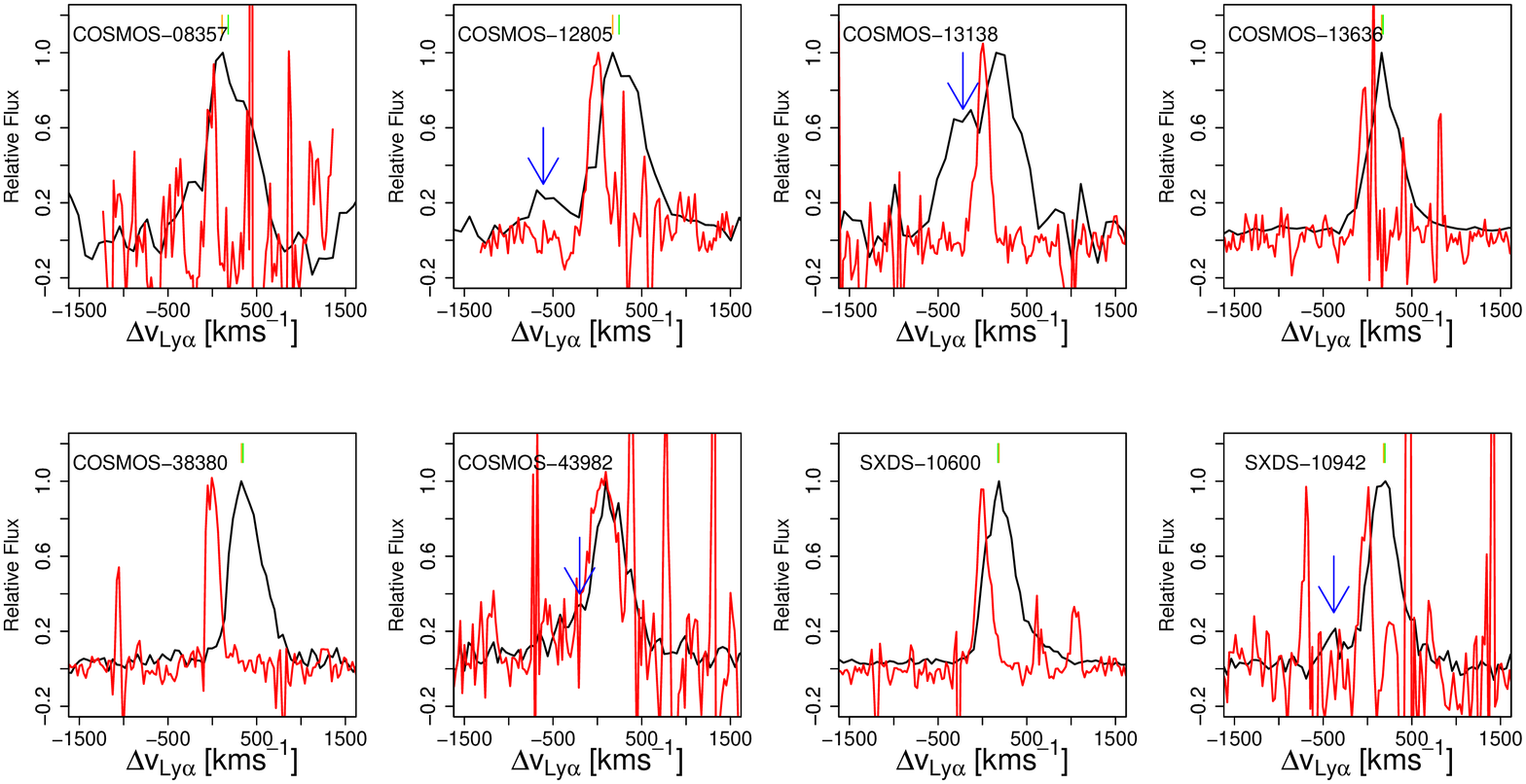} 
\caption[]
{
The upper two rows of panels show the Ly$\alpha$ lines (black) 
obtained with MagE and the corresponding H$\alpha$ lines (red), 
while the lower two rows of panels show the Ly$\alpha$ lines
obtained with LRIS (black) and the [OIII] lines (red). 
Blue arrows denote blue bumps. 
All spectra are scaled to the wavelength range from $-1500$ to $+1500$ km s$^{-1}$. 
Yellow and green segments indicate the peak flux positions 
derived from a symmetric Gaussian and a Monte Carlo technique, respectively. 
}
\end{figure*}

\begin{deluxetable*}{cccccccc}
\tabletypesize{\scriptsize}
\tablecolumns{8}
\tablewidth{0pt}
\tablecaption{Summary of the Observations \label{tab:summary_observations}}

\tablehead{
\colhead{Object} & \colhead{$\alpha$(J2000)} & \colhead{$\delta$(J2000)} & \colhead{EW(Ly$\alpha$)$_{\rm photo}$} & \colhead{$L(Ly\alpha)$} & \colhead{NIR obs.} & \colhead{opt. obs.} & \colhead{Source$^{a}$}\\
\colhead{} & \colhead{} & \colhead{} & \colhead{(\AA)} & \colhead{(10$^{42}$ erg s$^{-1}$)} &  \colhead{} & \colhead{} & \colhead{}\\
\colhead{(1)} & \colhead{(2)} & \colhead{(3)} & \colhead{(4)} &  \colhead{(5)} & \colhead{(6)} & \colhead{(7)} & \colhead{(8)}
}
\startdata
CDFS-3865 & 03:32:32.31  & -28:00:52.20 & $64\pm29$ & $29.8\pm4.9$ & NIRSPEC (J) & MagE & H13, N13 \\
 &  & &  & & MMIRS (HK)\\
\hline \\
CDFS-6482 & 03:32:49.34 & -27:59:52.35 & $76\pm52$ & $15.4\pm8.1$ & MMIRS (HK) & MagE & H13, N13\\
\hline \\
COSMOS-08501 & 10:01:16.80 & +02:05:36.26 & $280\pm30$ & $8.8\pm1.1$ & NIRSPEC (K) & MagE & N13\\
\hline \\
COSMOS-30679 & 10:00:29.81& +02:18:49.00  & $87\pm7$ & $8.5\pm0.7$ & NIRSPEC (H and K) & MagE & H13, N13\\
\hline \\
COSMOS-13636 & 09:59:59.38 & +02:08:38.36 & $73\pm5$ & $11.3\pm0.5$ & FMOS (H) & MagE and LRIS & H13, N13, S14\\
& &  &  & & NIRSPEC (K) & \\
\hline \\
COSMOS-43982$^{b}$ & 09:59:54.39 & +02:26:29.96 & $130\pm12$ & $11.0\pm0.5$ & MMIRS (HK) & MagE and LRIS & H13, N13, S14 \\
\hline \\
COSMOS-08357 & 09:59:59.07 & +02:05:31.60 & $47\pm8$ & $0.5\pm0.1$ & FMOS (H) & LRIS & S14, N15\\
\hline \\
COSMOS-12805 & 10:00:15.29 & +02:08:07.50 & $34\pm6$ & $2.6\pm0.3$ & FMOS (H) & LRIS & S14, N15\\ 
\hline \\
COSMOS-13138 & 10:00:02.61  & +02:08:24.50 & $40\pm10$ & $0.4\pm0.1$ & FMOS (H) & LRIS & S14, N15\\
\hline \\
COSMOS-38380 & 09:59:40.94  & +02:23:04.20 & $137\pm15$ & $2.6\pm0.3$ & FMOS (H) & LRIS & S14, N15 \\
\hline \\
SXDS-10600 & 02:17:46.09 & -06:57:05.00 &  $58\pm3$ & $1.9\pm0.1$ & FMOS (H)  & LRIS & S14, N15 \\
\hline \\
SXDS-10942 & 02:17:59.54 & -06:57:25.60 & $135\pm10$ & $0.3 \pm 0.0$ & FMOS (H) & LRIS & S14, N15
\enddata
\tablecomments{
(1) Object ID;
(2), (3) Right Ascension and Declination;
(4), (5) rest-frame Ly$\alpha$ EW and luminosity derived from narrow- and broadband photometry;
(6) Instruments and filters used for the NIR observations; 
(7) Instruments used for the optical observations;
and 
(8) Source of the information 
}
\tablenotetext{a}{
H13: \cite{hashimoto2013}; N13: \cite{nakajima2013}; S14: \cite{shibuya2014b}; N15: Nakajima et al (2015, in preparation)
}
\tablenotetext{b}{
AGN-like object
}
\end{deluxetable*}

\subsection{AGNs in the Sample} \label{subsec:agn}

The presence of AGNs in the MagE objects has been examined in \cite{hashimoto2013} and \cite{nakajima2013}, 
and those of the LRIS objects in \cite{shibuya2014b}. 

In short, for the MagE objects, we inspected it in three ways.  
We first compared the sky coordinates of the objects 
with those in very deep archival X-ray and radio catalogs. 
Then we checked for the presence of high ionization state lines 
such as {\sc Civ} $\lambda$ 1549 and He {\sc ii} $\lambda 1640$ lines in the spectra. 
Finally, we applied the BPT diagnostic diagram (\citealt{baldwin1981}) to the objects. 
No AGN activity is seen except for COSMOS-43982 
whose high [{\sc Nii}] /H$\alpha$ line ratio is consistent with that of an AGN. 

On the other hand, 
due to the lack of H$\alpha$ or [{\sc Nii}] $\lambda 6568$ data, 
we were only able to use the two forms of investigation for the LRIS objects. 
Of these, only 
COSMOS-43982 showed clear detection of 
the {\sc Civ} $\lambda$ 1549 line in its optical spectrum. 

In summary, we have ruled out AGN activity 
in all but COSMOS-43982.

\section{Observational Results} \label{sec:obs_results}

\subsection{Line Center and FWHM Measurements for Nebular Emission Lines} \label{subsec:obs_results1}

Line center (i.e., redshift) and FWHM measurements of nebular emission lines are crucial for a  
detailed modeling of the Ly$\alpha$ line, since they encode information on 
the intrinsic (i.e., before being affected by radiative transfer) 
Ly$\alpha$ redshift and FWHM. 
In order to obtain these parameters and their uncertainties, 
we apply a Monte Carlo technique as follows. 
First, for each line of each object, we measure the 1$\sigma$ 
noise of the local continuum. 
Then we create 10$^{3}$ fake spectra by perturbing the flux at each wavelength of the 
true spectrum by the measured 1$\sigma$ error (\citealt{kulas2012, chonis2013}). 
For each fake spectrum, 
the wavelength at the highest flux peak is adopted as the line center, 
and the wavelength range 
encompassing half the maximum flux is adopted as the FWHM. 
The standard deviation of the distribution of measurements from the 10$^{3}$ 
artificial spectra is adopted as the error on the line center and FWHM. 
When multiple lines are detected, we adopt a weighted mean value of them.
A summary of the measurements are listed in the columns 2 and 3 of 
Table \ref{tab:summary_obs_results}. 
All redshift (FWHM) values are corrected for the LSR motion (instrumental resolution). 
When the line is unresolved, 
the instrumental resolution is given as an upper limit.
The mean FWHM value for a sample of eight objects with a measurable velocity dispersion 
is FWHM(neb) $= 129\pm55$ km s$^{-1}$, which is smaller than that of LBGs,  
FWHM(neb) $= 200-250$ km s$^{-1}$ (\citealt{pettini2001, erb2006a, kulas2012}). 
This is consistent with the recent results by \cite{erb2014}, 
who have found that the median FWHM(neb) of 36 $z\sim2$ LAEs is 127 km s$^{-1}$. 
These results indicate that LAEs have smaller dynamical masses than LBGs.

\begin{deluxetable*}{ccccccccc}
\tabletypesize{\scriptsize}
\tablecolumns{9}
\tablewidth{0pt}
\tablecaption{Summary of the observed spectroscopic properties of the sample \label{tab:summary_obs_results}}

\tablehead{
\colhead{Object} & \colhead{$z_{\rm sys}$} & \colhead{FWHM(neb)} & \colhead{Blue Bump} & \colhead{$\Delta v_{\rm Ly\alpha , r}$} & \colhead{$\Delta v_{\rm Ly\alpha  , b}$} & \colhead{$\Delta v_{\rm peak}$} & \colhead{EW(Ly$\alpha$)$_{\rm spec}$} & \colhead{$S_{\rm w}$}\\
\colhead{} & \colhead{} & \colhead{(km s$^{-1}$)} & \colhead{} & \colhead{(km s$^{-1}$)} &  \colhead{(km s$^{-1}$)}  & \colhead{(km s$^{-1}$)} & \colhead{(\AA)} & \colhead{}\\
\colhead{(1)} & \colhead{(2)} & \colhead{(3)} & \colhead{(4)} &  \colhead{(5)} & \colhead{(6)} & \colhead{(7)} & \colhead{(8)} & \colhead{(9)} 
}
\startdata
CDFS-3865 & $2.17242\pm0.00016$  & $242\pm31$ & yes & $245\pm36$ & $-352\pm59$ &$597\pm67$  & $40\pm2$ & $8.8\pm0.3$\\
\hline \\
CDFS-6482 & $2.20490\pm0.00042$ & $99^{+66}_{-99}$ & no & $118\pm48$ & - & -  & $26\pm2$ & $6.6\pm1.7$\\
\hline\\
COSMOS-08501 & $2.16161\pm0.00042$ & $< 200$ & no & $82\pm40$ & - & -  & $10\pm1$ & $2.2\pm2.7$\\
\hline \\
COSMOS-30679 & $2.19725\pm0.00020$ & $92\pm45$ & no & $290\pm33$ & - & - &  $10\pm1$ & $3.1\pm1.0$\\
\hline \\
COSMOS-13636 (MagE) & $2.16075\pm0.00019$ & $73\pm5$ & no & $146\pm25$ & - & -  & $23\pm5$ & $5.3\pm1.0$\\
COSMOS-13636 (LRIS) & $2.16075\pm0.00019$ &  $73\pm5$ & no & $161\pm18$ & - & -  & $26\pm1$ & $6.2\pm0.5$\\
\hline \\
COSMOS-43982 (MagE) & $2.19267\pm0.00036$ & $325\pm36$ & yes & $117\pm53$ & $-297\pm57$ & $414\pm78$  & $24\pm17$ & $7.9\pm1.3$\\
COSMOS-43982 (LRIS) &  $2.19267\pm0.00036$ & $325\pm36$ & yes & $155\pm40$ & $-165\pm90$ & $320\pm98$  & $42\pm3$ & $-4.2\pm0.6$\\
\hline \\
COSMOS-08357 & $2.18053\pm0.00031$ & $<136$ & no & $106\pm71$ & - & -  & $19\pm3$ & $-4.2\pm7.9$\\
\hline \\	
COSMOS-12805 & $2.15887\pm0.00024$ & $110\pm16$  & yes & $171\pm25$ & $-605\pm114$ & $776\pm117$  & $24\pm1$ & $8.9\pm0.7$\\
\hline \\	
COSMOS-13138 & $2.17914\pm0.00012$ & $63\pm6$ & yes & $191\pm59$ & $-214\pm87$ & $405\pm105$ & $46\pm11$ & $-1.6\pm4.8$\\
\hline \\
COSMOS-38380 & $2.21245\pm0.00015$ & $99\pm9$ & no & $338\pm21$ & - & -  & $73\pm7$ & $2.5\pm0.8$\\
\hline \\
SXDS-10600 & $2.20922\pm0.00014$ & $55\pm28$ & no & $186\pm13$ & - & - &  $44\pm1$ & $11.7\pm0.2$\\
\hline \\
SXDS-10942 & $2.19574\pm0.00025$ & $<136$ & yes & $135\pm10$ & $-374\pm41$ & $556\pm66$ &  $94\pm10$ & $1.3\pm0.3$\\
\enddata
\tablecomments{
The symbol ``-'' indicates we have no measurement.
(1) Object ID;
(2) Systemic redshift derived from the weighted mean of the nebular emission redshifts;
(3) Weighted mean FWHM of nebular emission line;
(4) Presence of a blue bump emission in the Ly$\alpha$ profile;
(5) Velocity offset of the Ly$\alpha$ main red peak with respect to $z_{\rm sys}$; 
(6) Velocity offset of the Ly$\alpha$ blue-bump with respect to $z_{\rm sys}$; 
(7) Separation between $\Delta v_{\rm Ly\alpha , r}$ and $\Delta v_{\rm Ly\alpha , b}$;
(8) Rest-frame Ly$\alpha$ EW derived from spectroscopy;
and (9) Weighted skewness of the Ly$\alpha$ line.
}

\end{deluxetable*}

\subsection{Two Component {\rm[{\sc Oiii}]} Profiles} \label{subsec:obs_results2}

\begin{figure*}[]
\centering
\includegraphics[width=6cm]{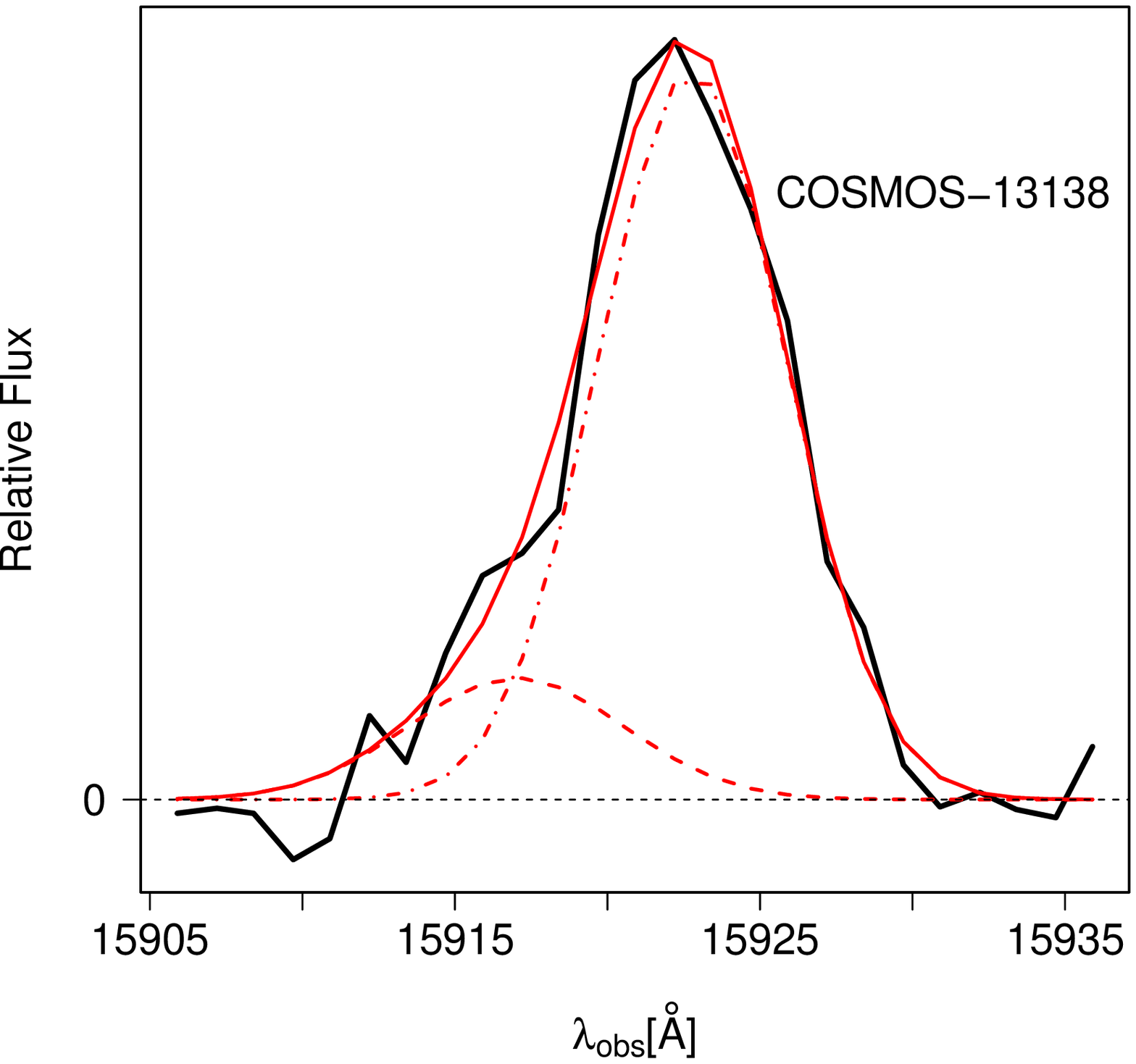} 
\includegraphics[width=6cm]{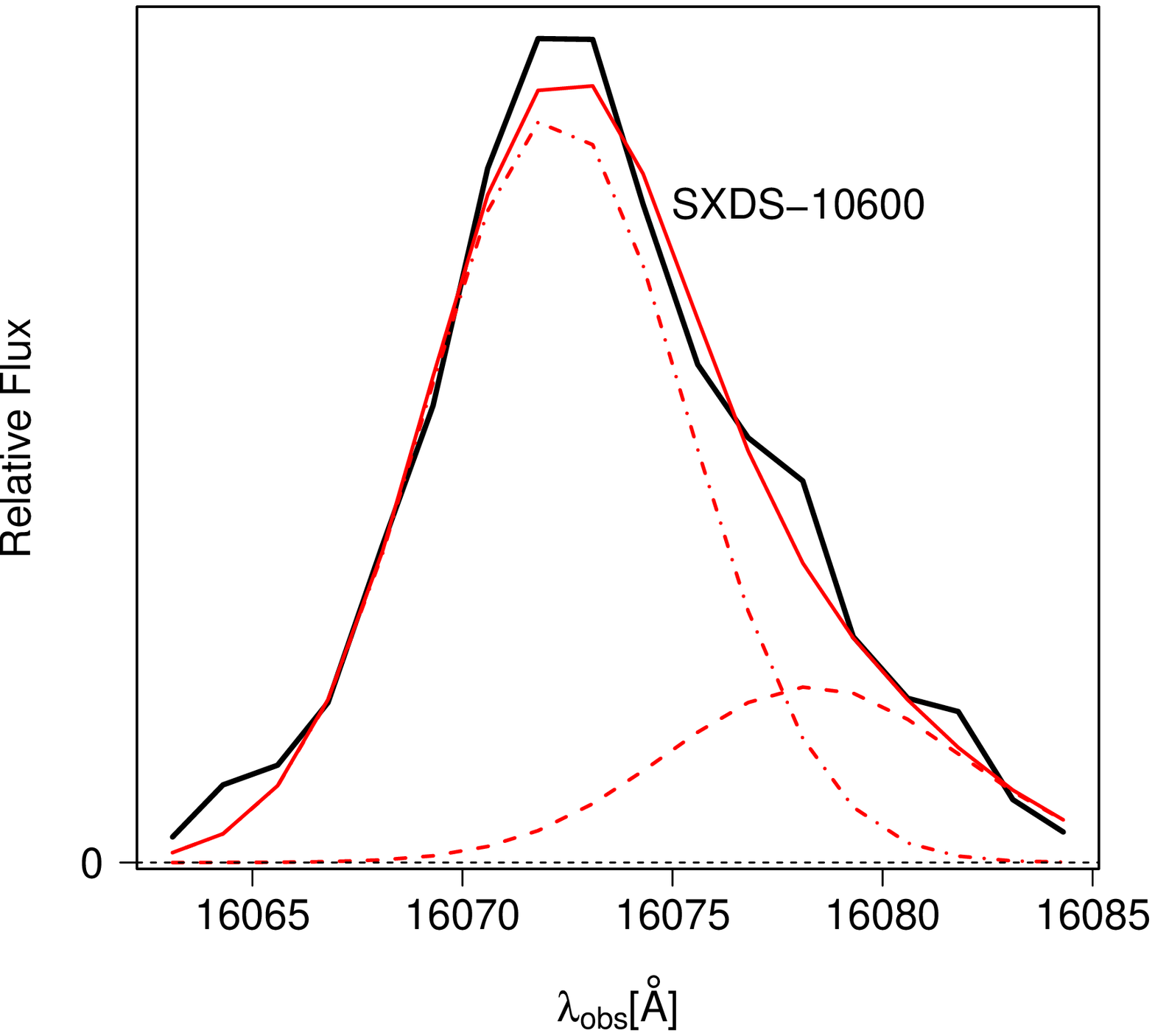} 
\caption[]
{
[{\sc Oiii}] $\lambda5007$ spectrum of COSMOS-13138 (left panel) and SXDS-10600 (right) taken by FMOS. 
The black solid lines are the observed spectra.  
There is a broad component in addition to a narrow component.  
The red solid lines denote the best-fit functions for the whole lines, 
and the red dashed lines represent those of the individual components.  
}
\label{fig:two-components}
\end{figure*}

Among the nebular emission lines we have obtained, 
while most objects show normal symmetric Gaussian profiles, 
COSMOS-13138 and SXDS-10600 
show an asymmetric [{\sc Oiii}] profile 
with a secondary blueshifted and redshifted component, respectively 
(see Figure \ref{fig:two-components}). 
Such a profile has been reported in various objects: 
both local and high-$z$ star-forming galaxies and ULIRGs 
(e.g., \citealt{shapiro2009, genzel2011, newman2012, soto2012}), 
a high-$z$ Oxygen-Two Blob ([{\sc Oii}] blob) \citep{harikane2014}, 
and a few Lyman-Alpha Blobs (LABs) \citep{yang2014}. 
However in LAEs, there has been no study which reports its presence.

Aforementioned studies apply a two Gaussian components fit 
with a narrow and broad components to the line. 
To examine the presence of two components, 
we also perform a fit with two Gaussians. 
We have six parameters: fluxes, line centers, and FWHMs for both components. 
We require that the widths of both components are larger than the spectral resolution, 
and that the broad component has a larger FWHM than the narrow component. 
Best fit parameters are determined through minimum $\chi^{2}$ realizations,
and the parameter range 
satisfying $\chi^{2}$ $\leq$ $\chi^{2}_{\rm min}$ + 1 
is adopted as the error, 
where $\chi^{2}_{\rm min}$ denotes the minimum $\chi^{2}$ value. 
The results are listed in Table \ref{tab:two-components}. 
For each object, both components are significantly detected with $\gtrsim4\sigma$, 
demonstrating that some fraction of LAEs have two-component line profiles.

The velocity offsets of the two components are 
$104\pm11$ km s$^{-1}$ (COSMOS-13138)
and 
$115\pm8$ km s$^{-1}$ (SXDS-10600), 
respectively.

The FWHM values of the broad component 
after correction for instrumental resolution are 
$70\pm50$ and $80\pm30$ km s$^{-1}$. 
These are much smaller than those of the star forming galaxies at 
$z\sim2$ (FWHM $ = 300 - 1000$ km s$^{-1}$, \citealt{genzel2011}), 
and slightly smaller than 
those of the [{\sc Oii}] blob (FWHM $ =120 - 130$ km s$^{-1}$) of  \cite{harikane2014} 
and the LABs (FWHM $ =100 - 280$ km s$^{-1}$) of \cite{yang2014}. 
Our small values exclude the possibility 
of the broad component  
originating from an AGN activity (cf., \citealt{osterbrock2006}) 
or a powerful outflow driven by a starburst (e.g., \citealt{shapiro2009, genzel2011, newman2012}) 
because in these cases, 
the FWHM of the broad component should be as large as $\sim300 - 1000$ km s$^{-1}$.
It is possible that the two component lines 
originate from two large star-forming regions (e.g., \citealt{harikane2014}) or mergers. 
As discussed in \cite{harikane2014}, 
the velocity offset of the two components, $\sim100$ km s$^{-1}$, 
may be due to a rotation of the objects.

\begin{deluxetable*}{ccccccc}
\tabletypesize{\scriptsize}
\tablecolumns{7}
\tablewidth{0pt}
\tablecaption{Summary of the Two-Component Fits for {\rm [{\sc Oiii}]} line \label{tab:two-components}   }

\tablehead{
\colhead{Object} & \colhead{$f_{\rm narrow}$} &\colhead{$f_{\rm broad}$} & \colhead{$z_{\rm narrow}$} & \colhead{$z_{\rm broad}$} & \colhead{FWHM$_{\rm narrow}$} & \colhead{FWHM$_{\rm broad}$}\\
\colhead{} & \colhead{(10$^{-17}$ erg s$^{-1}$ cm$^{-2}$)} & \colhead{(10$^{-17}$ erg s$^{-1}$ cm$^{-2}$)} & \colhead{} &\colhead{} &  \colhead{(km s$^{-1}$)}  & \colhead{(km s$^{-1}$)}\\
\colhead{(1)} & \colhead{(2)} & \colhead{(3)} & \colhead{(4)} &  \colhead{(5)} & \colhead{(6)} & \colhead{(7)} 
}
\startdata
COSMOS-13138 & $4.7\pm0.2$ & $0.9\pm0.2$ & $2.17931\pm0.00002$ & $2.17821\pm0.00012$ & $30\pm20$ & $70\pm50$\\
\hline \\
SXDS-10600 &  $14.9\pm0.6$ & $4.2\pm0.6$ & $2.20915\pm0.00002$ & $2.21038\pm0.00008$ & $20\pm30$ & $80\pm30$
\enddata
\tablecomments{
(1) Object ID;
(2), (3) Fluxes of the narrow ($f_{\rm narrow}$) and the broad ($f_{\rm broad}$) components. 
Note that the values are not corrected for the slit loss;
(4), (5) Redshifts of the narrow ($z_{\rm narrow}$) and the broad ($z_{\rm broad}$) components;
(6), (7) FWHM measurements of the narrow (FWHM$_{\rm narrow}$) and the broad (FWHM$_{\rm broad}$) components. 
}
\end{deluxetable*}

\subsection{Ly$\alpha$ Profile with a Blue Bump} \label{subsec:blue_bump}

While the majority of Ly$\alpha$ profiles are single-peaked (e.g., \citealt{shapley2003, steidel2010}), 
a fraction of Ly$\alpha$ profiles are known to be multiple-peaked (e.g., \citealt{rauch2008, yamada2012, kulas2012}). 
In particular, we shall refer to
a secondary small peak blueward of the systemic redshift as 
``the bluebump'' (see the case 2 profile in Figure 12 in \citealt{verhamme2006}). 
Theoretical studies have shown that the blue bump is a natural outcome 
of the radiative transfer in a low speed galactic outflow (e.g., \citealt{zheng2002}).

We consider a blue bump to be detected 
if there exists an excess emission blueward of the systemic redshift 
above $3\sigma$ noise of the local continuum. 
We detect a blue bump of five objects; 
the MagE ones of CDFS-3865 and COSMOS-43982, 
and the LRIS ones of COSMOS-12805, COSMOS-13138, 
COSMOS-43982, and SXDS-10942 (the column 4 of Table \ref{tab:summary_obs_results}). 
The position of the blue bump is designated by 
a blue arrow in Figure 1.

The frequency of blue-bump objects in the sample is $\sim40\%$ (5/12). 
There are four LAEs in the literature that have a blue bump: 
one among the two LAEs studied in \cite{mclinden2011} 
and all three LAEs studied in \cite{chonis2013}. 
For the total sample of 17 LAEs, 
the frequency is calculated to be $\sim50\%$ (9/17). 
Note that this is a lower limit due to the limited spectral resolution. 
On the other hand, 
\cite{kulas2012} have studied 18 $z\sim2-3$ LBGs with $z_{\rm sys}$ measurements 
which are preselected to have multiple-peaked Ly$\alpha$ profiles. 
They have argued that $\sim30\%$ of the parent sample are multiple-peaked 
and that 11 out of the 18 objects have a blue bump, 
indicating that the blue bump frequency in LBGs is $\sim20\%$ ($\sim30\% \times 11/18$). 
These results imply that the blue bump feature 
is slightly more common in LAEs than in LBGs 
although a larger sample observed at higher spectral resolution 
is needed for a definite conclusion.

\subsection{Ly$\alpha$ Velocity Properties} \label{subsec:obs_results3}

We derive three velocity offsets related to the Ly$\alpha$ line: 
the velocity offset of the main red peak of the Ly$\alpha$ line with respect to the systemic redshift, 
\begin{eqnarray}
\Delta v_{\rm Ly\alpha, r}
= c \frac{z_{{\rm Ly}\alpha, {\rm r}} - z_{\rm sys}}{1+z_{\rm sys}},
\end{eqnarray}
that of  the blue bump of the Ly$\alpha$ line with respect to the systemic redshift, if any, 
\begin{eqnarray}
\Delta v_{\rm Ly\alpha, b}
= c \frac{z_{{\rm Ly}\alpha, {\rm b}} - z_{\rm sys}}{1+z_{\rm sys}},
\end{eqnarray}
and that of the two peaks, 
\begin{eqnarray}
\Delta v_{\rm peak}
= \Delta v_{\rm Ly\alpha, r} - \Delta v_{\rm Ly\alpha, b},
\end{eqnarray}
where $z_{\rm sys}$, $z_{{\rm Ly}\alpha, {\rm r}}$, and $z_{{\rm Ly}\alpha, {\rm b}}$ 
represent the systemic redshift, the Ly$\alpha$ redshift of the main red peak, 
and that of the blue bump, respectively.

\subsubsection{Ly$\alpha$ Main Red Peak Velocity Offsets, $\Delta v_{\rm Ly\alpha, r}$} \label{subsec:obs_results3-1}

We estimate the $\Delta v_{\rm Ly\alpha, r}$ value 
using a Monte Carlo technique in a similar manner to that in \S \ref{subsec:obs_results1}. 
First, for each object, we measure the $1\sigma$ error in the Ly$\alpha$ spectrum 
set by the continuum level at the wavelength longer than $1216$\AA. 
Then we create 10$^{3}$ fake spectra converted to velocity space 
by simultaneously perturbing the flux at each wavelength and the systemic redshift 
listed in Table \ref{tab:summary_obs_results} by their $1\sigma$ errors. 
Finally, we measure the velocity at the highest flux peak.
The mean and the standard deviation value of the distribution of 10$^{3}$ measurements 
are adopted as the $\Delta v_{\rm Ly\alpha, r}$ and its error, respectively. 
The derived $\Delta v_{\rm Ly\alpha, r}$ values 
are listed in the column 5 of Table \ref{tab:summary_obs_results}, 
ranging from $82$ km s$^{-1}$ to $338$ km s$^{-1}$ 
with a mean value of $174\pm19$ km s$^{-1}$. 
In most cases, these values are consistent with those measured in 
\cite{hashimoto2013} and \cite{shibuya2014b} within $1\sigma$, however, 
they are not for COSMOS-08357 and COSMOS-12805. 
This is due to the fact that these studies have applied 
a symmetric/asymmetric profile fit to the Ly$\alpha$ line. 
In Figure 1, we show the two $\Delta v_{\rm Ly\alpha, r}$ values 
derived from the Monte Carlo and the profile fit technique as 
the orange and green line segments, respectively. 
For the sake of consistency in the definition of the $\Delta v_{\rm Ly\alpha, r}$ 
in the shell model (\citealt{verhamme2006, schaerer2011}), 
we adopt here the new measurements. 
We note that our discussion is unchanged even if we adopt the previous $\Delta v_{\rm Ly\alpha, r}$ values.

The $\Delta v_{\rm Ly\alpha, r}$ value has been measured in more than 60 LAEs 
(\citealt{mclinden2011, finkelstein2011b, hashimoto2013, 
guaita2013, chonis2013, shibuya2014b, song2014, erb2014}). 
These studies have shown that LAEs at $z\sim2-3$ have a mean 
$\Delta v_{\rm Ly\alpha, r}$ of $\simeq200$ km s$^{-1}$, 
which is significantly smaller than that of LBGs at a similar redshift, 
$\Delta v_{\rm Ly\alpha, r} \simeq 400$ km s$^{-1}$ (e.g., \citealt{steidel2010, rakic2011, kulas2012}).
The left panel of Figure \ref{fig:histogram_velocity} represents the histogram of $\Delta v_{\rm Ly\alpha, r}$ 
for the 12 LAEs (14 spectra) studied in this study and 18 LBGs given by \cite{kulas2012}. 
We carry out the Kolmogorov-Smirnov (K-S) test for the two populations. 
The resultant probability is $10^{-6}$, indicating that $\Delta v_{\rm Ly\alpha, r}$ 
is definitively different between LAEs and LBGs.

\subsubsection{Ly$\alpha$ Blue Bump Velocity Offsets, $\Delta v_{\rm Ly\alpha, b}$} \label{subsec:obs_results3-2}

For each detected blue bump in \S \ref{subsec:blue_bump}, 
we measure $\Delta v_{\rm Ly\alpha, b}$ value 
in the same manner as for $\Delta v_{\rm Ly\alpha, r}$. 
We obtain 
$\Delta v_{\rm Ly\alpha, b} = $ $-352\pm59$ km s$^{-1}$ (CDFS-3865), 
$-297\pm57$ km s$^{-1}$ (MagE-COSMOS-43982), 
$-605\pm114$ km s$^{-1}$ (COSMOS-12805),
$-214\pm87$ km s$^{-1}$ (COSMOS-13138), 
$-165\pm90$  km s$^{-1}$ (LRIS-COSMOS-43982), 
and $-374\pm41$ km s$^{-1}$ (SXDS-10942) 
as listed in the column 6 of Table \ref{tab:summary_obs_results}. 
We have obtained two different measurements for COSMOS-43982 
due to the spectral resolution effect, however, they are consistent with each other 
within $1\sigma$ (see also \S \ref{subsubsec:4-3-3}).
We combine our $\Delta v_{\rm Ly\alpha, b}$ measurements 
with those in the four aforementioned LAEs with a blue bump 
to construct a large sample of LAEs with a blue bump 
consisting of 9 objects (10 spectra): 
one from \cite{mclinden2011} with $\Delta v_{\rm Ly\alpha, b} = -454$  km s$^{-1}$ 
and three from \cite{chonis2013} with $\Delta v_{\rm Ly\alpha, b} = -127, -250$, 
and $-236$ km s$^{-1}$. 
The mean $\Delta v_{\rm Ly\alpha, b}$ value of the  large sample is 
$\Delta v_{\rm Ly\alpha, b} = -316\pm45$ km s$^{-1}$, 
which is consistent with that of 11 LBGs with a blue bump, 
$\Delta v_{\rm Ly\alpha, b} = -367\pm46$ km s$^{-1}$ (\citealt{kulas2012}). 
We calculate the K-S probability to be 0.3901, 
indicating that LAEs' $\Delta v_{\rm Ly\alpha, b}$ values 
are comparable to LBGs'. 
The middle panel of Figure \ref{fig:histogram_velocity} shows 
the $\Delta v_{\rm Ly\alpha, b}$ distribution for the LAE and LBG samples.

We check if our conclusion remains unchanged 
even if the spectral resolution effect is taken into account. 
The sample by \cite{kulas2012} has been obtained 
with three settings: 300-line grating, 400-, and 600-line grisms,
corresponding to a spectral resolution of 
$R\sim600, 800,$ and $1300$, respectively. 
We compare the mean $\Delta v_{\rm Ly\alpha, b}$ value of 
our four LAEs taken by LRIS ($R\sim1100$) 
and that of six LBGs with a blue bump obtained at a similar resolution ($R\sim1300$). 
The resultant  mean $\Delta v_{\rm Ly\alpha, b}$ values for LAEs and LBGs are  
$-340\pm99$ and $-356\pm70$ km s$^{-1}$, respectively, 
and the K-S probability is 0.9238. 
Thus, we obtain the same conclusion.

\subsubsection{Velocity Offsets Between the Main Red Peak and the Blue Bump, $\Delta v_{\rm peak}$} \label{subsec:obs_results3-3}

\begin{figure*}[]
 \centering
 \includegraphics[width=5cm]{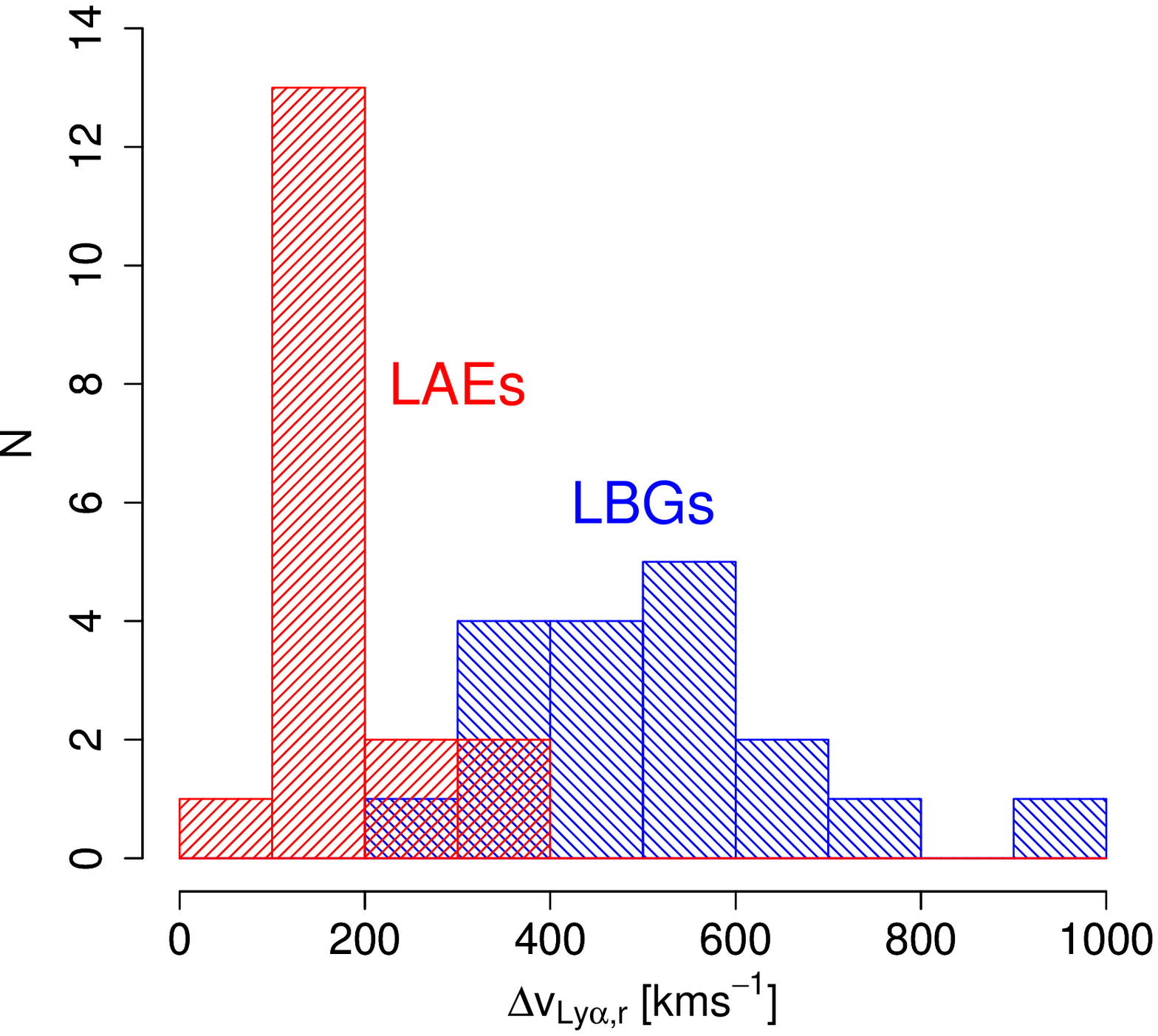} 
 \includegraphics[width=5cm]{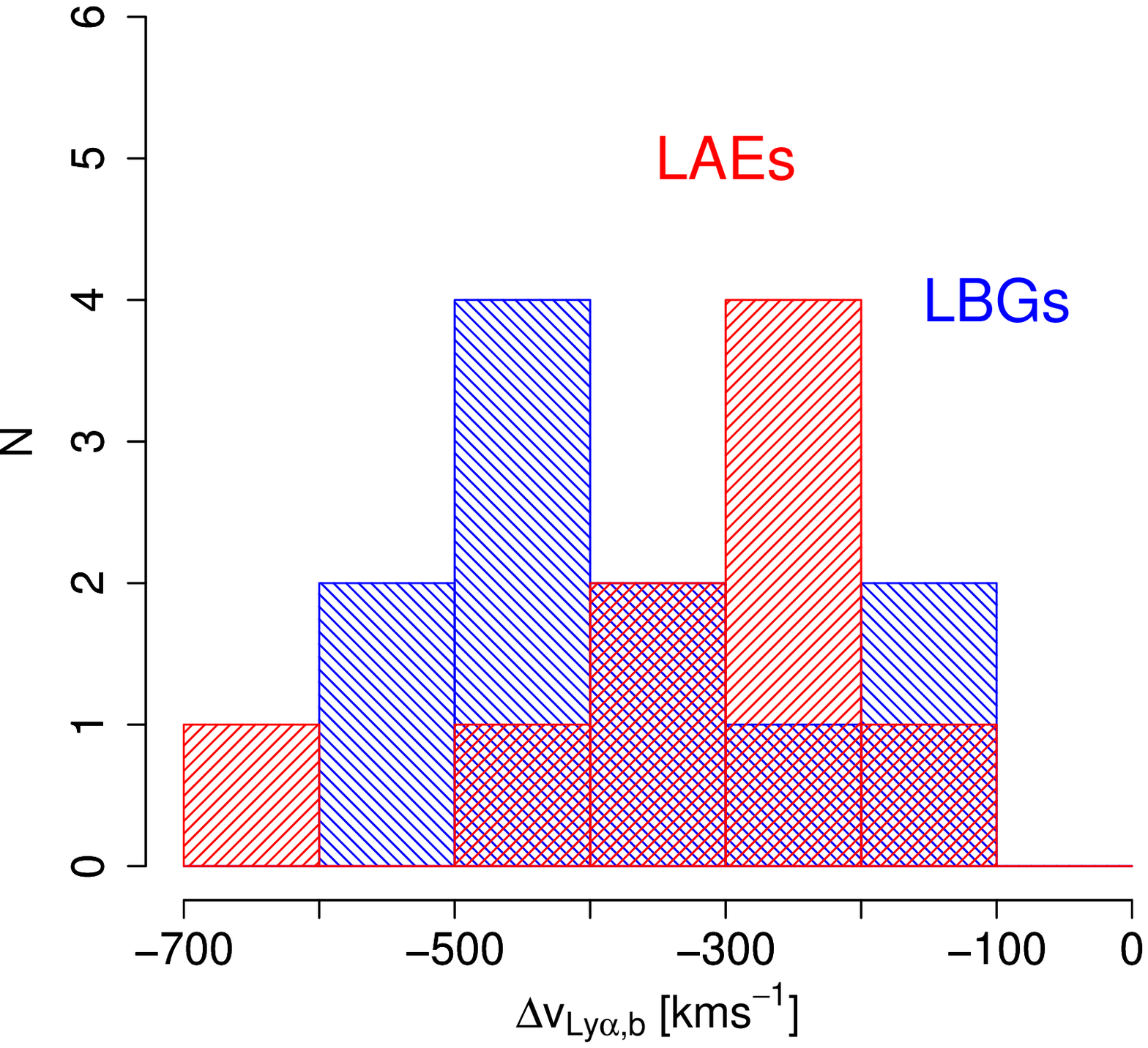} 
 \includegraphics[width=5cm]{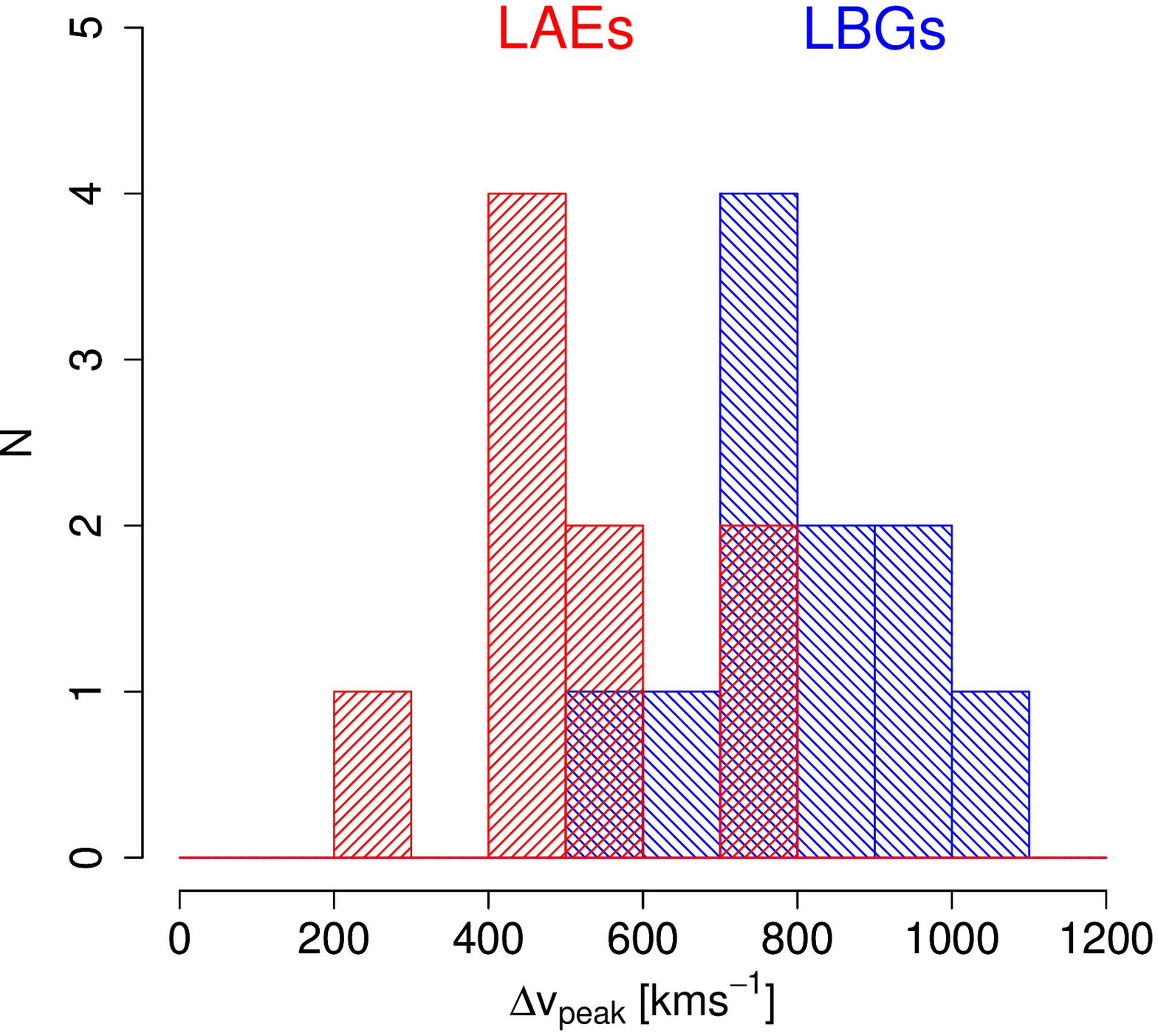} 
\caption[]
{
Histograms of $\Delta v_{\rm Ly\alpha, r}$ (left panel), $\Delta v_{\rm Ly\alpha, b}$ (middle), and $\Delta v_{\rm peak}$ (right) 
for the LAEs studied in this paper and the literature (\citealt{mclinden2011, chonis2013}) and LBGs given by \cite{kulas2012}. 
}
\label{fig:histogram_velocity}
\end{figure*}

Finally, 
for each of the spectra with a blue bump, we measure 
the velocity offset between the red and blue peaks: 
$\Delta v_{\rm peak}$ = $597\pm67$ km s$^{-1}$ (CDFS-3865), 
$414\pm78$ km s$^{-1}$ (MagE-COSMOS-43982), 
$776\pm117$ km s$^{-1}$ (COSMOS-12805), 
$405\pm105$ km s$^{-1}$ (COSMOS-13138), 
$320\pm98$ km s$^{-1}$ (LRIS-COSMOS-43982),
and $556\pm66$ km s$^{-1}$ (SXDS-10942), 
as listed in the column 7 of Table \ref{tab:summary_obs_results}. 
In order to make a large sample with $\Delta v_{\rm peak}$ measured, 
we utilize again the four LAEs with the blue bump from the literature: 
one LAE studied in \cite{mclinden2011} with $\Delta v_{\rm peak} = 796$ km s$^{-1}$  
and three LAEs studied in \cite{chonis2013} with $\Delta v_{\rm peak} = 300, 425$, and $415$ km s$^{-1}$.  
The mean value of the nine objects (ten spectra) is $\Delta v_{\rm peak} = 500\pm56$ km s$^{-1}$, 
which is significantly smaller than the value derived for 11 LBGs with a blue bump, 
$\Delta v_{\rm peak} = 801\pm41$ km s$^{-1}$ (Group I in \citealt{kulas2012}). 
The K-S probability is calculated to be $0.00636$, indicating that LAEs and LBGs 
have distinctive $\Delta v_{\rm peak}$ values. 
See the right panel of Figure \ref{fig:histogram_velocity} for 
their distributions.

We examine the spectral resolution effect 
exactly the same manner as in \S \ref{subsec:obs_results3-2}. 
The mean $\Delta v_{\rm peak}$ value of 
the four LAEs taken by LRIS ($R\sim1100$) 
and that of the six LBGs with a blue bump obtained at a similar spectral resolution ($R\sim1300$) 
are $\Delta v_{\rm peak} = 514\pm100$ and $778\pm59$ km s$^{-1}$, respectively.
In conjunction with the K-S probability, 0.09524, 
we conclude that LAEs have a significantly smaller $\Delta v_{\rm peak}$ value than 
that  of LBG even at the same spectral resolution. 
Our finding is recently supported by \cite{henry2015} and \cite{yang2015}, 
who have examined Ly$\alpha$ velocity properties and their relations to the Ly$\alpha$ escape fraction 
for local galaxies called ``Green Peas'' galaxies (\citealt{cardamone2009}). 
They have found that the Ly$\alpha$ escape fraction is higher for objects with smaller $\Delta v_{\rm peak}$.

In summary, we have derived three Ly$\alpha$ velocity offsets, $\Delta v_{\rm Ly\alpha, r}$, $\Delta v_{\rm Ly\alpha, b}$, 
and $\Delta v_{\rm peak}$. 
While we need a larger sample of objects with a blue bump for a definite conclusion, 
we find that LAEs have a smaller (comparable) 
$\Delta v_{\rm Ly\alpha, r}$ ($\Delta v_{\rm Ly\alpha, b}$) value 
relative to LBGs, which makes their $\Delta v_{\rm peak}$ value also smaller than that of LBGs.

\begin{deluxetable*}{cccccc}
\tabletypesize{\scriptsize}
\tablecolumns{6}
\tablewidth{0pt}
\tablecaption{Summary of the Other Physical Properties \label{tab:summary_other_prop}}

\tablehead{
\colhead{Object} & \colhead{$\Delta v_{\rm abs}$} & \colhead{log($M_{*}$)} & \colhead{$E(B-V)_{*}$} & \colhead{merger} & \colhead{$\epsilon$} \\
\colhead{} & \colhead{} & \colhead{} & \colhead{} &  \colhead{(pair, $CAS$)} & \colhead{} \\
\colhead{} & \colhead{(km s$^{-1}$)} & \colhead{($M_{\odot}$)} & \colhead{(mag.)} &\colhead{} & \colhead{}\\
\colhead{(1)} & \colhead{(2)} & \colhead{(3)} & \colhead{(4)} &  \colhead{(5)} & \colhead{(6)}
}
\startdata
CDFS-3865 & ($-102\pm65$) & $9.50^{+0.01}_{-0.03}$ & $0.14^{+0.00}_{-0.00}$ & -, -  & - 
\\
\hline 
CDFS-6482 & ($-102\pm65$) & $9.80^{+0.06}_{-0.05}$ & $0.15^{+0.02}_{-0.02}$ & -, -  & - 
\\
\hline 
COSMOS-08501 & - & $7.84^{+1.21}_{-0.27}$ & $0.08^{+0.04}_{-0.08}$ & no, -  & - 
\\
\hline 
COSMOS-30679 & ($-102\pm65$) & $9.74^{+0.26}_{-0.52}$ & $0.24^{+0.04}_{-0.04}$ & no, -  & $0.24$ 
\\ 
\hline 
COSMOS-13636 & $-130\pm70$ ($-102\pm65$) & $9.12^{+0.13}_{-0.14}$ & $0.18^{+0.01}_{-0.01}$ & yes, no  & - 
\\
\hline
COSMOS-43982 & - & $10.80^{+0.01}_{-0.06}$ & $0.40^{+0.00}_{-0.00}$ & no, no  & 0.49
\\
\hline
COSMOS-08357 &- & $9.21^{+0.28}_{-0.40}$ & $0.14^{+0.05}_{-0.05}$ & no, -  & - 
\\
\hline 
COSMOS-12805 & $-170\pm50$ & $9.44^{+0.13}_{-0.17}$ & $0.16^{+0.02}_{-0.02}$ & yes, -  & - 
\\
\hline 
COSMOS-13138 & - & $9.48^{+0.22}_{-0.20}$ & $0.19^{+0.04}_{-0.04}$ & -, - & - 
\\
\hline
COSMOS0-38380 & - & $10.06^{+0.06}_{-0.11}$ & $0.13^{+0.02}_{-0.01}$ & no, no  & $0.34$ 
\\
\hline
SXDS-10600 & $-260\pm60$ & $9.46^{+0.05}_{-0.04}$ & $0.05^{+0.00}_{-0.01}$ & -, -  & - 
\\
\hline
SXDS-10942 &- & $7.73^{+0.11}_{-0.08}$ & $0.04^{+0.02}_{-0.02}$ & -, - & - 
\enddata
\tablecomments{
The symbol ``-'' indicates no measurement.
(1) Object ID;
(2) Mean velocity offset of LIS absorption lines with respect to $z_{\rm sys}$; 
(3) Stellar mass estimated from SED fitting; 
(4) Stellar dust extinction estimated from SED fitting; 
(5) Presence of merger examined via close-pair method and $CAS$ system studied in \cite{shibuya2014a}; 
(6) Ellipticity defined as $\epsilon = 1 - b/a$, where $a$ and $b$ are the major and minor axes, respectively. 
}
\end{deluxetable*}

\subsection{Other Physical Quantities} \label{subsec:other_quantities}

In this section, we describe other physical quantities related to with this work. 
We describe 
metal absorption line properties in \S \ref{subsec:obs_results4-1}, 
SED fitting properties in \S \ref{subsec:obs_results4-2}, 
and morphological properties in \S \ref{subsec:obs_results4-3}. 

\subsubsection{Metal Absorption Line Properties} \label{subsec:obs_results4-1}

Low ionization state (LIS) metal absorption lines encode information 
on cold neutral gas in galaxies. 
The mean blueshift of LIS absorption lines 
with respect to the systemic velocity, $\Delta v_{\rm abs}$, 
gives the average speed of the galactic outflow (e.g., \citealt{pettini2001, shapley2003, martin2005}). 
In the following sections, 
we compare the $\Delta v_{\rm abs}$ values  of our LAE sample 
with the results from Ly$\alpha$ radiation transfer fitting.

\cite{shibuya2014b} have detected several LIS absorption lines  
in a few narrowband-selected LAEs on the individual basis. 
The derived mean blue shifts are 
$\Delta v_{\rm abs} = $ $-130\pm70$ km s$^{-1}$ (COSMOS-13636), 
$-170\pm50$ km s$^{-1}$ (COSMOS-12805), 
and $-260\pm60$ km s$^{-1}$ (SXDS-10600). 
Additionally, \cite{hashimoto2013} have detected several LIS absorption lines 
in a stacked spectrum of four LAEs: CDFS-3865, CDFS-6482, COSMOS-13636, 
and COSMOS-30679. 
The mean blueshift of the LIS metal absorption lines is 
$\Delta v_{\rm abs} = $ $-102\pm65$ km s$^{-1}$. 
These values are listed in the column 2 of Table \ref{tab:summary_other_prop}. 

\subsubsection{SED Fitting Properties} \label{subsec:obs_results4-2}

In this study, we utilize SED fitting results of the sample, in particular, 
stellar dust extinction, $E(B-V)_{*}$, 
and stellar mass, $M_{*}$. 
In the following sections, 
we compare the $E(B-V)_{*}$ values 
with the results from Ly$\alpha$ radiation transfer fitting, 
and investigate the correlation between the Ly$\alpha$ profile trends and $M_{*}$

SED fitting results for the MagE (LRIS) objects 
have been presented in \cite{hashimoto2013} and \cite{nakajima2013} (\citealt{shibuya2014b}). 
For the detail procedure of the fitting, 
we refer the reader to \cite{ono2010b,ono2010a}.
The derived $E(B-V)_{*}$ and $M_{*}$ values are listed 
in the columns 3 and 4 in Table \ref{tab:summary_other_prop}.  
The former range from $E(B-V)_{\rm *} = 0.04$ to  $0.40$ with a mean value of $E(B-V)_{\rm *} = 0.16$, 
and the latter from log $M_{*} / M_{\odot}$ = 7.7 to  10.8  
with a mean of log $M_{*} / M_{\odot}$ = 9.3, respectively.

\subsubsection{Morphological Properties} \label{subsec:obs_results4-3}

In the following sections, 
we use three morphological properties 
studied for $z\sim2.2$ LAEs in \cite{shibuya2014a}: 
the presence of a merger, the spatial offset between 
Ly$\alpha$ and stellar-continuum emission peaks, 
$\delta_{\rm Ly\alpha}$,  and the ellipticity. 
\cite{shibuya2014a} have utilized $I_{\rm 814}$ and $H_{\rm 160}$ data 
taken with ACS and WFC3 on $HST$ 
to examine rest-frame UV and optical morphologies, respectively. 
Among the objects presented in this study, 
the rest-frame UV images of the eight COSMOS objects 
have been investigated in \cite{shibuya2014a}.

The presence of a merger has been examined with two methods: 
the close-pair method(e.g., \citealt{lefevre2000, law2012}) 
and the morphological index method, 
especially $CAS$ system (\citealt{abraham1996, conselice2000}). 
In \cite{shibuya2014a}, the former method has been applied  
to objects with $I_{\rm 814} < 26.5$, which is the case for all the COSMOS 
objects presented in this study except for COSMOS-13138. 
The result is that two objects, COSMOS-13636 and COSMOS-12805, 
are mergers, while the remaining seven are not. 
On the other hand, 
the latter method has been done for objects with $I_{\rm 814} < 25.0$ 
and a half light radius, $r_{\rm e}$, larger than $0.''09$. 
The reason why \cite{shibuya2014a} have limited the sample for the latter method is 
to obtain reliable values of the indices.
This is the case for three COSMOS objects presented in this study, 
COSMOS-13636, COSMOS-43982, and COSMOS-38380. 
The result is that none of the three is a merger. 
The two results for COSMOS-13636 are not consistent with each other 
because we have used two different methods. 
Thus, among the eight COSMOS objects, 
COSMOS-13636 and COSMOS-12805 may be a merger 
(the column 5 of Table \ref{tab:summary_other_prop}).

The Ly$\alpha$ spatial offset, $\delta_{\rm Ly\alpha}$, 
has been examined by performing source detections 
with {\tt SExtractor} for Subaru NB387 and $HST$ $I_{\rm 814}$ images. 
While compact objects with a symmetric UV light profile 
tend to have a small $\delta_{\rm Ly\alpha}$ value, 
objects with an asymmetric, disturbed UV light profile 
likely to have a large $\delta_{\rm Ly\alpha}$ value (e.g., \citealt{jiang2013, shibuya2014a}).
Thus, this quantity could be a useful tracer 
of the {\sc Hi} gas stability around the galaxy. 
The value is reliably obtained for the objects with $I_{\rm 814} < 26.5$ and NB387 $<24.5$, 
where the typical positional error in $I_{\rm 814}$ (NB387) is less than $0.''02$ ($0.''3$). 
For the eight COSMOS objects in this study, 
none has a significant Ly$\alpha$ spatial offset 
larger than the typical error of the $\delta_{\rm Ly\alpha}$, $\sim 0.''36$.

The ellipticity, $\epsilon = 1 - a/b$, where $a$ and $b$ are the major and minor axes, 
is a useful indicator of the galactic disk inclination. 
In \cite{shibuya2014a}, 
this has been measured using {\tt GALFIT} software (\citealt{peng2002}) for the objects 
with $I_{\rm 814}$ $<$ 25.0 
and $r_{\rm e}$ larger than the typical PSF size. 
The former criterion, corresponding to $S/N = 30$ detection, 
is needed for the reliable ellipticity measurements 
(e.g.,\citealt{mosleh2012, ono2013}).
Only three objects, 
COSMOS-30679, COSMOS-38380, and COSMOS-43982, 
satisfy these criteria. 
The resultant ellipticity values are $\epsilon = $ 0.24 (COSMOS-30679), 
0.34 (COSMOS-38380), and 0.49 (AGN-COSMOS-43982), respectively 
(the column 6 of Table \ref{tab:summary_other_prop}).

\section{Ly$\alpha$ radiative transfer model and fitting procedure}\label{sec:model_fitting}

\subsection{A Library of Synthetic Spectra}\label{subsec:model}

The library of synthetic Ly$\alpha$ spectra used in this study 
has been described in \cite{schaerer2011}. 
Ly$\alpha$ radiation transfer has been computed with McLya \citep{verhamme2006}
through spherically symmetric expanding shells of 
homogeneous and isothermal neutral hydrogen gas.  
The shell is describe by four parameters: 
\begin{itemize}
\item the radial expansion velocity, $V_{\rm exp}$,   
\item the neutral hydrogen column density along the line of sight, 
$N_{\rm HI}$, 
\item the Doppler parameter, $b$, describing the thermal and turbulent 
motion in the shell, 
\item and the dust absorption optical depth at the Ly$\alpha$ wavelength, $\tau_{\rm a}$, 
related to the gas dust extinction by $E(B-V)_{\rm gas}\approx (0.06...0.11)\tau_{\rm a}$, 
where the lower and higher values in the parenthesis correspond to the 
attenuation law for starbursts (\citealt{calzetti2000}) 
and the Galactic extinction law (\citealt{seaton1979}), respectively. 
\end{itemize}
The Ly$\alpha$ source is located at the center of the shell. 
The intrinsic (i.e., before being affected by the radiative transfer effect)
spectrum is a Gaussian Ly$\alpha$ line plus a flat continuum, 
and is characterized by two parameters : 
\begin{itemize} 
\item the Ly$\alpha$ equivalent width, EW$_{\rm int}$(Ly$\alpha$), 
\item and the full width at half maximum, FWHM$_{\rm int}$(Ly$\alpha$). 
\end{itemize}
For a comparison with the observed data, 
each rest-frame model has been shifted using the systemic redshift $z_{\rm sys}$ values 
listed in Table \ref{tab:summary_obs_results}. 
To reflect the $z_{\rm sys}$ uncertainty, we have allowed the observed Ly$\alpha$ spectra 
to shift relative to the velocity zero point within the error. 
Thus, combinations of seven free parameters are fitted to the data.

This library of Ly$\alpha$ spectra has been successfully used to reproduce 
various observed Ly$\alpha$ line profiles of $z > 3$ LBGs, 
from strong emission to broad absorption
\citep{verhamme2008, schaerer2008, dessauges-zavadsky2010, vanzella2010, lidman2012}.

\subsection{Fitting of Observed Spectra}\label{subsec:fitting}

To perform a statistical comparison between the observed and modeled Ly$\alpha$ line profiles, 
we calculate the $\chi^2$ values for each of the possible combinations of the parameters 
for each galaxy (cf., \citealt{chonis2013}). 
Since model spectra are normalized and at an infinite spectral resolution, 
two steps are needed before the $\chi^2$ calculation. 
First, we normalize the observed spectra 
using the continuum level estimated at wavelengths longer than 1216\AA. 
Second, each model Ly$\alpha$ spectrum has been convolved with 
a Gaussian whose FWHM corresponds to spectral resolutions:
\begin{equation}
{\rm FWHM} = c / R, 
\end{equation}   
where $c$ is the speed of light.

We note that our fitting technique 
gives exactly the same statistical weight 
to all data points of the continuum and the Ly$\alpha$ line. 
Finally for the sake of consistency, 
for each object we calculate $\chi^{2}$ in the wavelength range 
from $-3 \times {\rm FWHM_{obs} (Ly\alpha)}$ to  $+3 \times {\rm FWHM_{obs}(Ly\alpha)}$ 
around the Ly$\alpha$ line center.

\begin{figure*}[]
\centering
 \includegraphics[width=18cm]{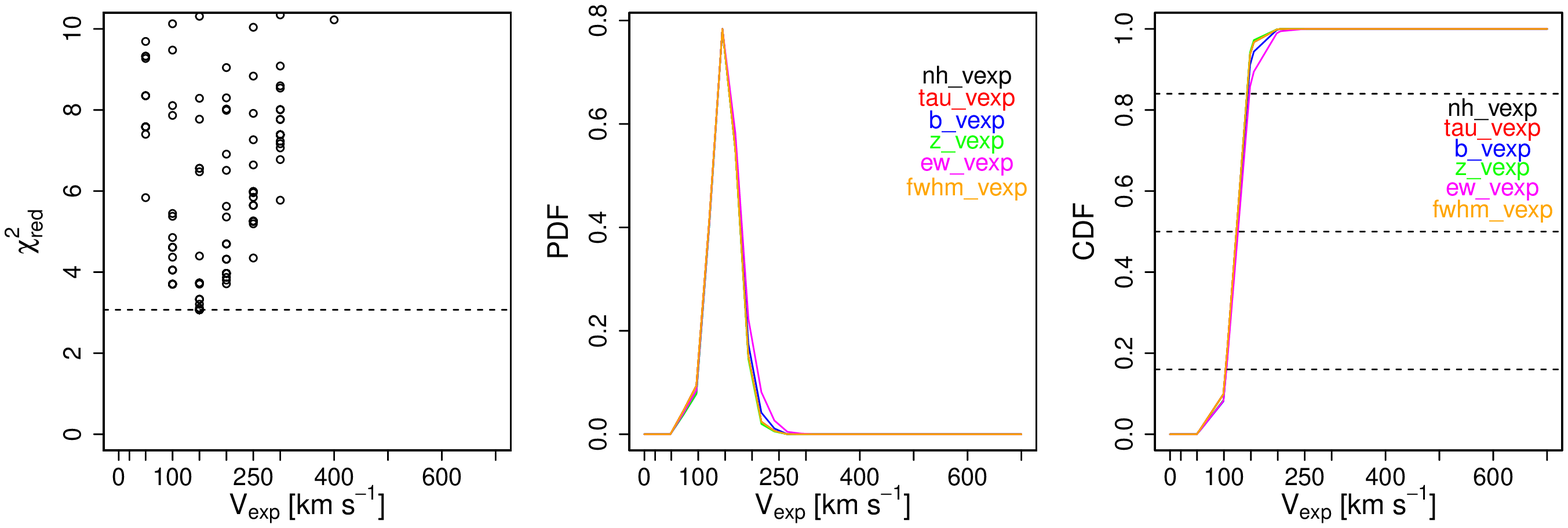}
 \includegraphics[width=18cm]{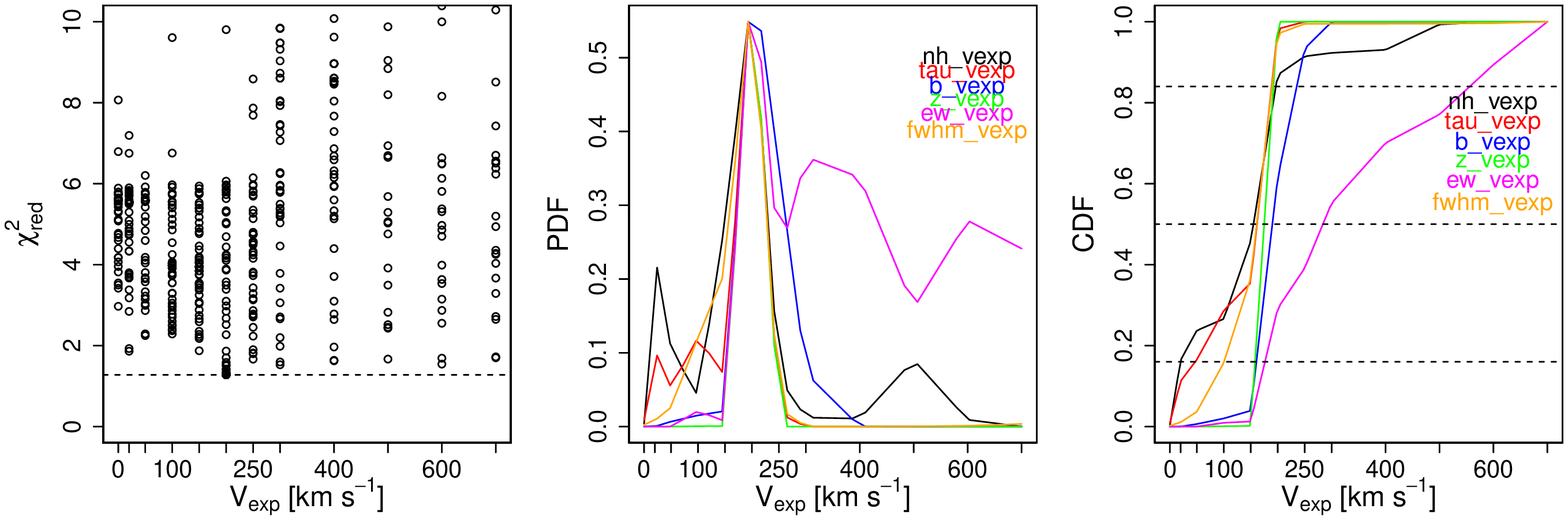}
\caption[]
{
Examples of reduced $\chi$$^{2}$ values (left panels),  
converted probability density functions (PDF) (middle panels), 
and cumulative density functions (CDF) (right panels), 
for the parameter $V_{\rm exp}$. 
The upper (lower) panels are for CDFS-3865 (COSMOS-08357).
}
\label{fig:example_chi2-pdf-cdf}
\end{figure*}

In Figure \ref{fig:example_chi2-pdf-cdf},  
we demonstrate how the best fit, and its associated errors, 
are found using  $\chi^{2}$ values. 
To do this, examples of the fit to $V_{\rm exp}$ 
are shown for well and poorly constrained objects. 
In the left panels of this figure, one can see a broad range of $V_{\rm exp}$ values 
with low reduced $\chi^{2}$ for COSMOS-08357 whose Ly$\alpha$ S/N ratio is $\sim11$, 
in comparison to CDFS-3865 with a Ly$\alpha$ S/N of $\sim 98$. 
To measure median and $1\sigma$ values, 
we convert $\chi^2$ values into probabilities using the formula, $p \propto$ exp($-\chi^{2}/2$) 
for each five 2D parameter set ($V_{\rm exp}$ vs. $N_{\rm HI}$,  $V_{\rm exp}$ vs. 
$\tau_{\rm a}$,  $V_{\rm exp}$ vs. $b$,  $V_{\rm exp}$ vs. FWHM$_{\rm int}$(Ly$\alpha$), 
and $V_{\rm exp}$ vs. EW$_{\rm int}$(Ly$\alpha$)). 
After normalizing them so that the total probability is unity, 
we draw a probability (PDF) and a cumulative density function (CDF)
as shown in the middle and the right panels, respectively. 
Finally, we adopt the values where the CDF value satisfying CDF $=0.50, 0.16,$ and $0.84$ 
as the median and $\pm1 \sigma$, respectively. 
Performing this for each five 2D parameter set 
results in five median and $\pm1 \sigma$ values. 
As can be seen, 
all the five median and $\pm 1\sigma$ values are 
consistent with each other for CDFS-3865, 
whereas those are not for COSMOS-08357.
In the latter case, we adopt the average of the five median and $\pm1 \sigma$ values.

\subsection{Results}\label{subsec:fitting_results}

We show the reproduced Ly$\alpha$ profiles (\S \ref{subsubsec:4-3-1}), 
describe the derived parameters (\S \ref{subsubsec:4-3-2}), 
and examine the influence of spectral resolution on the results (\S \ref{subsubsec:4-3-3}).

\subsubsection{Fitted Profiles}\label{subsubsec:4-3-1}

\begin{figure*}[]
\centering
 \includegraphics[width=19cm]{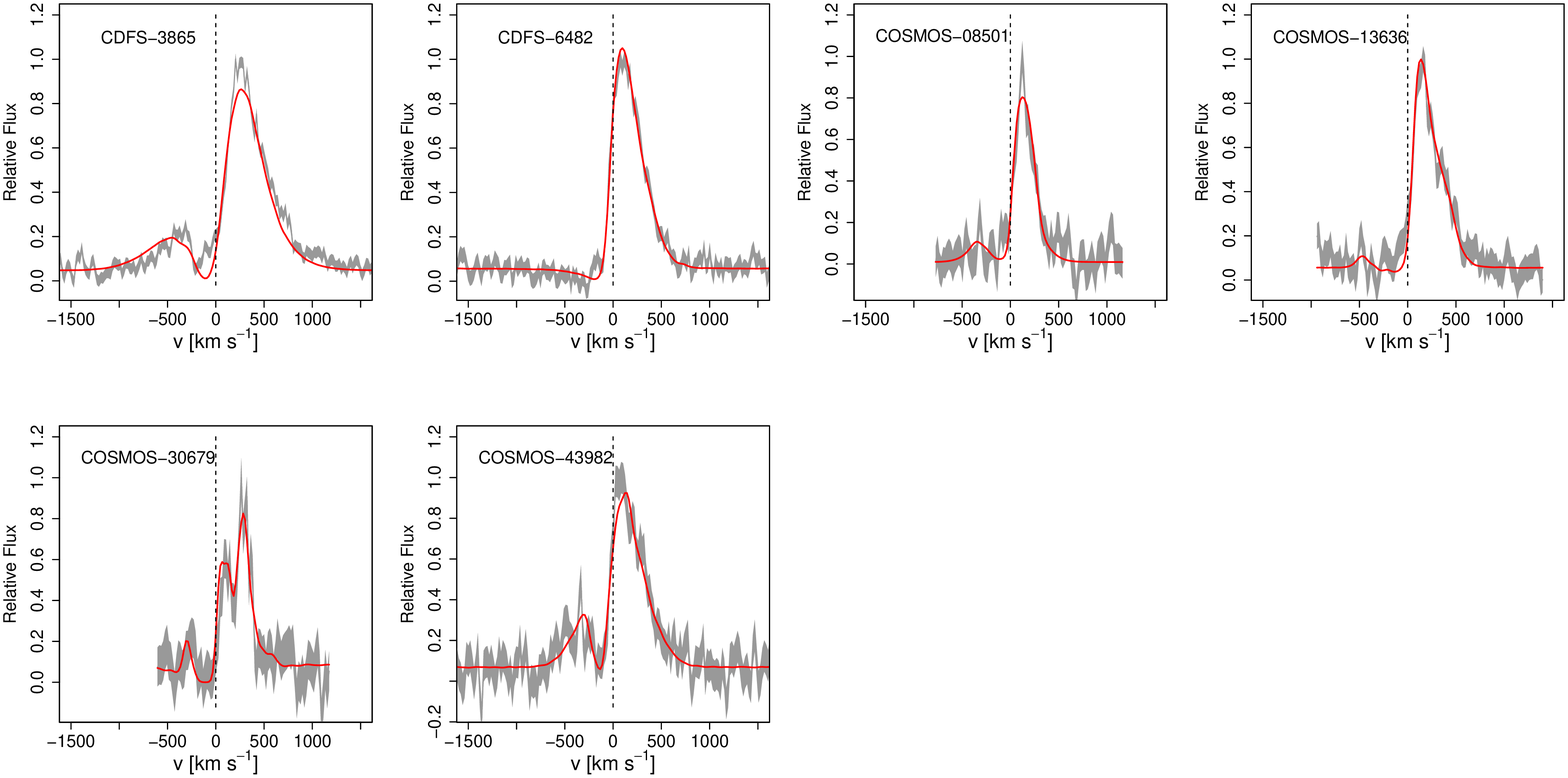}
 \includegraphics[width=19cm]{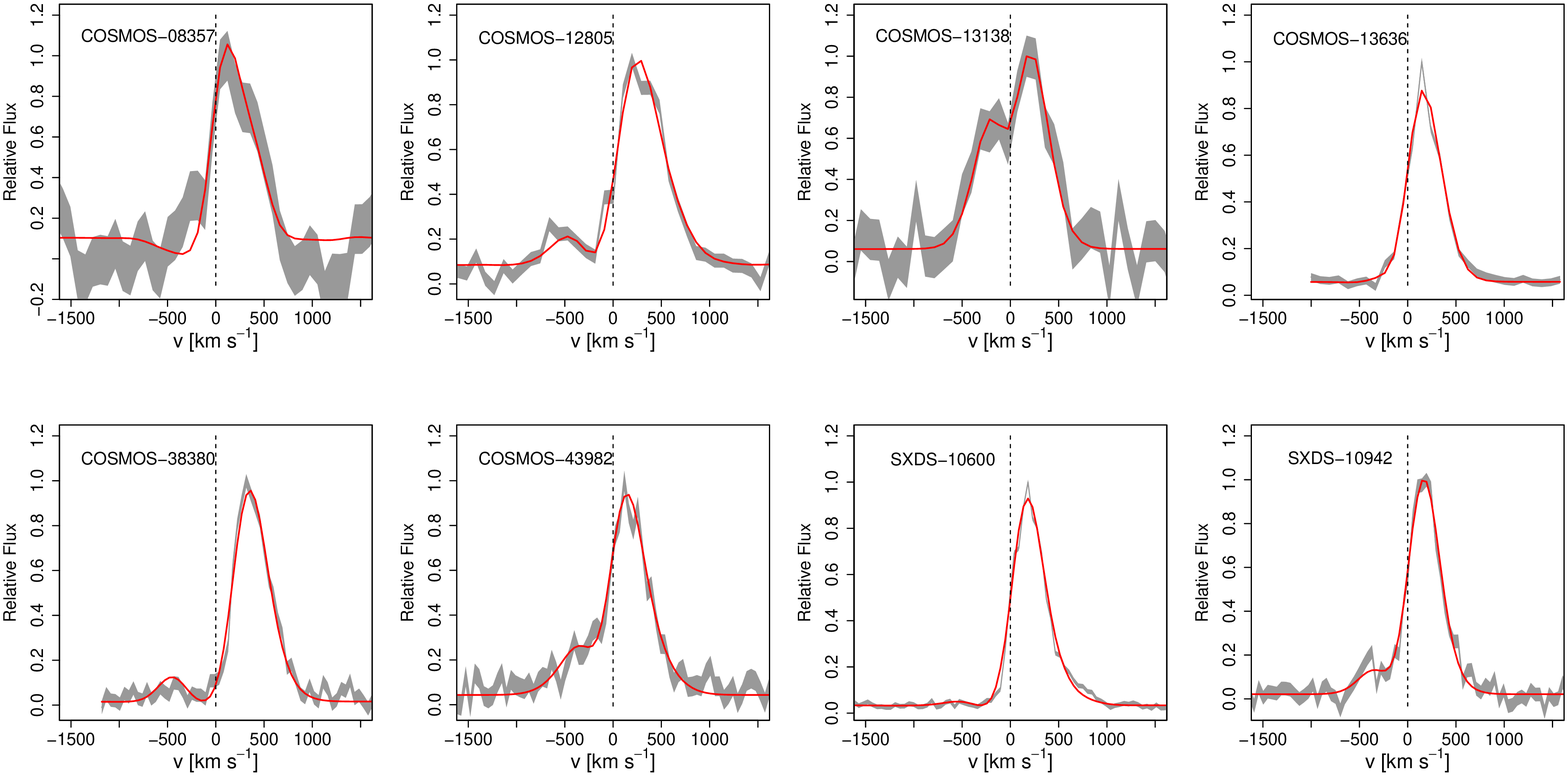}
\caption[]
{
The upper two rows of panels show  
the reproduced Ly$\alpha$ line profiles (red) on top of the observed ones (grey) for the MagE objects, 
while the lower two rows of panes are those for the LRIS objects. 
The gray region denotes the 1$\sigma$ range of the observed spectrum.
All spectra are scaled to the wavelength range from $-1500$ to $+1500$ km s$^{-1}$. 
}
\label{fig:fitted_spectrum}
\end{figure*}

Figure \ref{fig:fitted_spectrum} shows the best fit model spectra with the observed ones. 
All the Ly$\alpha$ profiles are quite well reproduced by the model, 
which seems to differ from the previous studies by
\cite{kulas2012} and \cite{chonis2013}. 
These authors have had difficulty reproducing their Ly$\alpha$ profiles, 
especially the position and the flux of the blue bump. 
This might be due to model differences. 
These two studies have utilized the uniform expanding shell model 
constructed by \cite{zheng2002} and \cite{kollmeier2010}. 
There are three major differences between the models (c.f., \citealt{chonis2013}). 
First, in addition to the three common parameters, $V_{\rm exp}$, $N_{\rm HI}$, and $b$, 
the model used in this study also includes an additional one for dust absorption. 
Second, the grid points and the physical range of parameters are different. 
The model by \cite{zheng2002} and \cite{kollmeier2010} 
has four values for each parameter:  
$V_{\rm exp}$ = 50, 100, 200, 300 km s$^{-1}$, 
log ($N_{\rm HI}$) = 17, 18, 19, 20.3 cm$^{-2}$, 
and $b$ = 20, 40, 80, 120 km s$^{-1}$, whereas 
the model used in this study has 12 $V_{\rm exp}$, 13 $N_{\rm HI}$, and 5 $b$ values
spanning wider physical ranges. 
Finally, the intrinsic spectrum of the previous models 
is a monochromatic Ly$\alpha$ line, 
while we model a Gaussian Ly$\alpha$ plus a continuum. 
As we show in \S \ref{subsubsec:4-3-2} and later sections, 
we infer that the key to better reproducing the blue bump 
is to  assume the Ly$\alpha$ profile to be a (broad) Gaussian.

\subsubsection{Derived Parameters}\label{subsubsec:4-3-2}

The best fit parameters are summarized in Table \ref{tab:summary_model}. 
We describe the mean values of the derived parameters, 
and systematically compare them with those of LBGs 
modeled by the same code (\citealt{verhamme2008, schaerer2008, dessauges-zavadsky2010}).
For the parameter FWHM$_{\rm int}$(Ly$\alpha$), 
we examine the mean values of two subsamples, 
objects with a blue bump and those without. 
We have checked that there is no significant difference 
between the two subsamples in the other parameters.

The mean $V_{\rm exp}$ value of the LAEs is 
$148 \pm 14$ km s$^{-1}$, 
which is comparable to that of LBGs, 
$131 \pm 25$ km s$^{-1}$. 
This strongly disfavors the hypothesis that 
the small $\Delta v_{\rm Ly\alpha}$ of LAEs is due to their 
large outflow velocity.

The most interesting parameter, $N_{\rm HI}$, 
ranges from log($N_{\rm HI}$) = 16.0 to 19.7 cm$^{-2}$, 
with a mean value of $18.4\pm0.4$ cm$^{-2}$,
which is more than one order of magnitude smaller than the typical 
log($N_{\rm HI}$) value of LBGs, $19.8\pm0.2$ (cm$^{-2}$).

The mean values of $\tau_{\rm a}$ and $b$ are 
$0.9\pm0.2$ and $37\pm10$ km s$^{-1}$,  
respectively, 
both of which are comparable to those of LBGs, 
$0.8\pm0.1$ and $28\pm5$ km s$^{-1}$.

FWHM$_{\rm int}$(Ly$\alpha$) values range 
from FWHM$_{\rm int}$(Ly$\alpha$) =  50 to 847 km s$^{-1}$.  
The mean values for the whole sample, the non blue bump sample, and 
the blue bump sample, are 354, 169, and 602 km s$^{-1}$, respectively. 
This shows that the blue bump objects have significantly larger FWHM$_{\rm int}$(Ly$\alpha$) 
than that found in the non blue bump objects. 
This trend is similar to \cite{verhamme2008}; 
They have found that most LBGs with a single peaked Ly$\alpha$ profile 
are best fitted with moderate values of FWHM$_{\rm int}$(Ly$\alpha$), $\sim200$ km s$^{-1}$,  
whereas the best fit FWHM$_{\rm int}$(Ly$\alpha$) values for 
two LBGs with a blue bump are greater than $500$ km s$^{-1}$. 
These results support our claim that large FWHM$_{\rm int}$(Ly$\alpha$) 
helps fitting the blue bump. 
We investigate if there are any observational trends for the blue bump objects, 
and discuss possible mechanisms for the blue bump objects 
to have large FWHM$_{\rm int}$(Ly$\alpha$) 
 in \S \ref{subsec:explanation_blue_bump}.

Since starburst activities that produce Ly$\alpha$ photons 
should be similar between LAEs and LBGs, 
we expect comparable EW$_{\rm int}$(Ly$\alpha$) values 
for these two galaxy populations. 
The result is that the mean EW$_{\rm int}$(Ly$\alpha$) value of LAEs, 
$65\pm18$ \AA, 
is somewhat smaller than that of LBGs, 
$107\pm25$ \AA.

In summary, 
the model parameter $N_{\rm HI}$ derived in LAEs is 
more than one order of magnitude smaller than that of LBGs, 
whereas the remaining parameters are consistent within $1 \sigma$ between LAEs and LBGs.

\subsubsection{Influence of Spectral Resolution on the Fitting Procedure}\label{subsubsec:4-3-3}

To investigate the influence of spectral resolution on the fitting results, 
we compare the best fit parameters of the two objects 
observed with the two spectrographs, COSMOS-13636 and COSMOS-43982. 
As can be seen in Table \ref{tab:summary_model}, 
the two fitting results of COSMOS-43982 are consistent with each other, 
whereas those of COSMOS-13636 are not, 
possibly owing to the large difference in the best-fit reduced $\chi^{2}$, 1.1 and 6.2. 

Taking a closer look into these two fits, 
we see that the extremely small 1 $\sigma$ noise in the flux of LRIS-COSMOS-13636 
could be a key reason for its high $\chi^{2}$ value. 
On the other hand, the modeled spectrum seems to be over-smoothed, 
leading us to infer its Ly$\alpha$ line resolution is under-estimated. 
Indeed, it is known that the spectral resolution for a given line can be higher than the canonical value. 
A combination of these factors would naturally cause the large resultant $\chi^{2}$ value, 
and the discrepancy between the different best-fit parameters at two resolutions.

\begin{deluxetable*}{ccccccccc}
\tabletypesize{\scriptsize}
\tablecolumns{9}
\tablewidth{0pt}
\tablecaption{Summary of the Ly$\alpha$ fitting for the sample \label{tab:summary_model}}

\tablehead{
\colhead{Object} & \colhead{$\chi^{2}_{\rm red}$} &\colhead{$V_{\rm exp}$}  & \colhead{log($N_{\rm HI}$)} & \colhead{$\tau_{a}$}& \colhead{$b$} & \colhead{FWHM(Ly$\alpha$)$_{\rm int.}$} & \colhead{EW(Ly$\alpha$)$_{\rm int.}$}\\
\colhead{} & \colhead{} & \colhead{(km s$^{-1}$)} & \colhead{(cm$^{-2}$)} & \colhead{} &  \colhead{(km s$^{-1}$)} & \colhead{(km s$^{-1}$)} & \colhead{(\AA)}\\
\colhead{(1)} & \colhead{(2)} & \colhead{(3)} & \colhead{(4)} &  \colhead{(5)} & \colhead{(6)} & \colhead{(7)} & \colhead{(8)}
}
\startdata
CDFS-3865 & 3.1 & $120^{+21}_{-14}$ & $19.5^{+0.1}_{-0.1}$ & $0.0^{+0.0}_{-0.0}$ & $15^{+13}_{-5}$ & $846^{+106}_{-97}$ & $35^{+7}_{-7}$ \\
\hline \\
CDFS-6482  & 1.3 & $ 177^{+18}_{-18}$ & $19.2^{+0.1}_{-0.1}$ & $0.12^{+0.04}_{-0.08}$ & $10^{+8}_{-0}$ & $271^{+38}_{-29}$ & $28^{+0}_{-7}$ \\
\hline \\
COSMOS-08501  & 1.3 & $167^{+286}_{-106}$ & $18.7^{+0.5}_{-1.1}$ & $1.56^{+1.54}_{-1.07}$ & $13^{+14}_{-3}$ & $252^{+240}_{-134}$ & $14^{+7}_{-7}$ \\
\hline \\
COSMOS-30679 & 1.0 & $127^{+14}_{-21}$ & $19.5^{+0.1}_{-0.1}$ & $1.43^{+1.03}_{-0.53}$ & $29^{+8}_{-8}$ & $50^{+38}_{-0}$ & $39^{+1}_{-11}$ \\
 \hline \\
COSMOS-13636 (MagE) & 1.1 & $226^{+14}_{-21}$ & $16.0^{+0.0}_{-0.0}$ & $0.12^{+0.14}_{-0.08}$ & $121^{+27}_{-27}$ & $256^{+101}_{-58}$ & $28^{+0}_{-7}$ \\
COSMOS-13636 (LIRS) & 6.2 & $127^{+14}_{-21}$ & $18.8^{+0.2}_{-0.2}$ & $0.08^{+0.08}_{-0.04}$ & $30^{+8}_{-6}$ & $127^{+19}_{-19}$  & $28^{+0}_{-7}$\\
 \hline \\
COSMOS-43982 (MagE) & 1.0 & $141^{+88}_{-57}$ & $18.2^{+0.6}_{-1.6}$ & $1.15^{+1.54}_{-0.89}$ & $12^{+11}_{-2}$ & $544^{+120}_{-134}$ & $28^{+7}_{-11}$ \\
COSMOS-43982 (LRIS) & 1.4 & $138^{+85}_{-71}$ & $18.1^{+0.4}_{-1.5}$ & $0.02^{+0.22}_{-0.02}$ & $13^{+8}_{-3}$ & $621^{+53}_{-86}$ & $42^{+7}_{-7}$ \\
 \hline \\
COSMOS-08357 & 1.3 & $170^{+25}_{-42}$ & $19.7^{+0.1}_{-0.6}$ & $2.24^{+1.25}_{-0.95}$ & $19^{+14}_{-9}$ & $74^{+82}_{-24}$ & $85^{+42}_{-35}$ \\
  \hline \\	
COSMOS-12805 & 3.1 & $177^{+18}_{-21}$ & $19.2^{+0.1}_{-0.1}$ & $1.73^{+0.71}_{-0.38}$ & $10^{+18}_{-0}$ & $645^{+38}_{-38}$ & $42^{+7}_{-0}$ \\
 \hline \\	
COSMOS-13138  & 1.5 & $21^{+481}_{-21}$ & $18.8^{+0.4}_{-0.7}$ & $1.13^{+1.45}_{-0.89}$ & $15^{+14}_{-5}$ & $501^{+144}_{-144}$ & $64^{+11}_{-11}$ \\
  \hline \\
COSMOS-38380 & 2.1 & $127^{+14}_{-21}$ & $19.7^{+0.1}_{-0.1}$ & $0.69^{+0.28}_{-0.32}$ & $60^{+14}_{-14}$ & $99^{+9}_{-9}$ & $276^{+14}_{-21}$ \\
  \hline \\
SXDS-10600  & 6.2 & $226^{+14}_{-21}$ & $16.0^{+0.0}_{-0.0}$ & $1.74^{+0.20}_{-0.16}$ & $121^{+27}_{-27}$ & $223^{+19}_{-19}$ & $113^{+7}_{-7}$ \\
  \hline \\
SXDS-10942 & 1.6 & $131^{+32}_{-35}$ & $16.0^{+0.0}_{-0.0}$ & $0.12^{+0.08}_{-0.08}$ & $60^{+14}_{-14}$ & $453^{+82}_{-67}$ & $85^{+14}_{-7}$ \\
\enddata
\tablecomments{
(1) Object ID;
(2) Reduced $\chi^{2}$ value of the fitting calculated as $\chi^{2}_{\rm red} = \chi^{2} / (N-M)$, 
where $N$ and $M$ denote the number of data points and the degree of freedom, respectively; 
(3) $-$ (8) 
Best fit values of the radial expansion velocity, the column density of the neutral Hydrogen, 
the dust absorption optical depth, the Doppler parameter, the intrinsic Ly$\alpha$ FWHM, 
and the intrinsic Ly$\alpha$ EW, respectively.
}
\end{deluxetable*}

\subsection{Degeneracy among Parameters} \label{subsec:degeneracy}

In this subsection, 
we investigate degeneracies among the model parameters 
to understand how they affect our determination of the best fit parameters. 
First we describe possible degeneracies and 
then statistically examine them using 2D $\chi^{2}$ values.

It is possible that parameters $\tau$ and EW$_{\rm int}$(Ly$\alpha$) are degenerated  
as an observed profile can be reproduced equivalently well 
either assuming a weak intrinsic line with low dust extinction, 
or a strong intrinsic line with high dust extinction. 
There would also be a degeneracy between $b$ and FWHM$_{\rm int}$(Ly$\alpha$) 
in the sense that both broaden the line profile.
Furthermore, when there is a blue bump in the profile, 
we need either a high $b$ or a low $V_{\rm exp}$ 
to reproduce it.

Figures 13 - 15 in the Appendix are 2D parameter grid maps for CDFS-3865 
with the grey dots showing the entire grids. 
We use these maps and $\chi^{2}$ values 
to examine the actual degeneracies among the parameters. 
If there is a degeneracy between two parameters, 
the $\chi^{2}$ contour would be tilted and elongated. 
The blue grids in these figures show 
those satisfying $\Delta \chi^{2} \leq 6.17$ 
above the raw minimum $\chi^{2}$ designated by the white dots, 
i.e., the 3 $\sigma$ uncertainty in the parameter set (\citealt{press1992}). 
Thanks to the number of data points given by high spectral resolutions, 
and the relatively coarse grids, 
even the 3 $\sigma$ uncertainty is converged into one grid. 
This indicates that there is no degeneracy that affects our determination 
of the best fit. 
We have checked that this is also true for the rest of the sample in this study. 
Thus, we conclude that the systematic uncertainties among the parameters 
due to the degeneracies are small, 
and thus do not affect our discussions.

\subsection{Comparison between Observation and Model} \label{subsec:comp_obs_model}

In order to examine if the best fit parameters 
are reasonable, we compare the derived parameters 
with the observables.

\subsubsection{$|\Delta v_{\rm abs}|$ vs. $V_{\rm exp}$} \label{subsubsec:comp_vexp}
  
\begin{figure}[]
\centering
 \includegraphics[width=8cm]{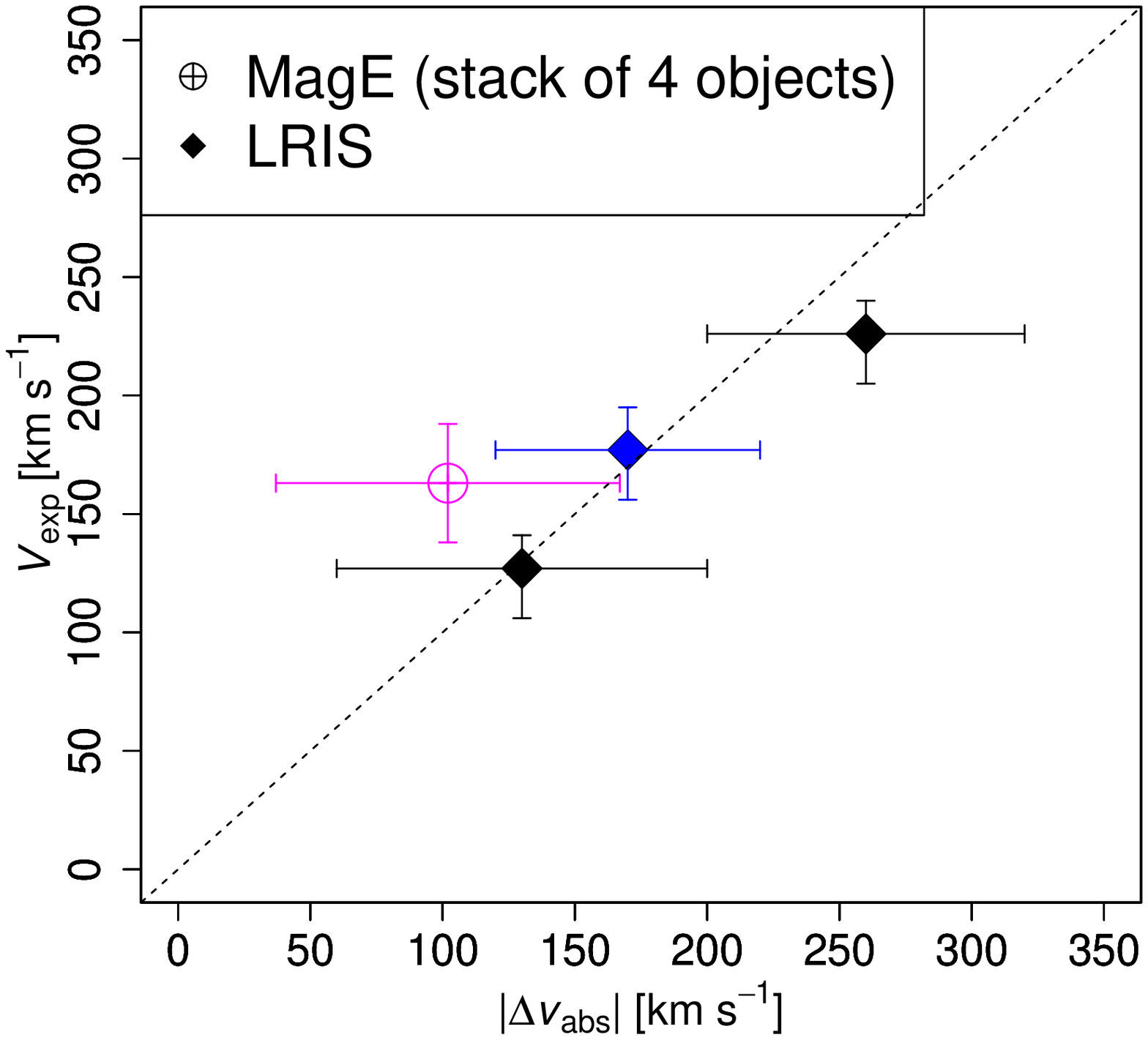}
\caption[]
{
$V_{\rm exp}$  plotted against $|\Delta v_{\rm abs}|$. 
The diamonds are the three LRIS objects 
whose LIS absorption lines are detected in the individual spectrum (\citealt{shibuya2014b}). 
Blue color shows an object with a blue bump in the Ly$\alpha$ profile, 
while black color denotes those without. 
The magenta circle is the stacked spectrum of the four MagE LAEs, 
CDFS-3865, CDFS-6482, COSMOS-13636, and COSMOS-30679 (\citealt{hashimoto2013}). 
}
\label{fig:comp_vexp}
\end{figure}

As stated in \S \ref{subsec:obs_results4-1}, 
several LIS absorption lines have been detected 
in individual spectra of COSMOS-12805, COSMOS-13636, 
and SXDS-10600 (\citealt{shibuya2014b}), 
and in a stacked spectrum of four LAEs, CDFS-3865, CDFS-6482, COSMOS-13636, 
and COSMOS-30679 (\citealt{hashimoto2013}). 
The measured blueshift of LIS absorption lines with respect to the systemic, 
$\Delta v_{\rm abs}$, is listed in Table \ref{tab:summary_other_prop}. 
Figure \ref{fig:comp_vexp} shows a comparison between 
$|\Delta v_{\rm abs}|$ and 
the best-fit expansion velocity, $V_{\rm exp}$. 
For the stacked spectrum, 
we plot the mean $V_{\rm exp}$ value of the four LAEs, $163\pm25$ km s$^{-1}$. 
While there are only four data points, 
$|\Delta v_{\rm abs}|$ and $V_{\rm exp}$ are in excellent agreement with each other.

\subsubsection{E(B-V)$_{*}$ vs. $\tau_{\rm a}$ } \label{subsubsec:comp_dust}

\begin{figure}[]
\centering
 \includegraphics[width=8cm]{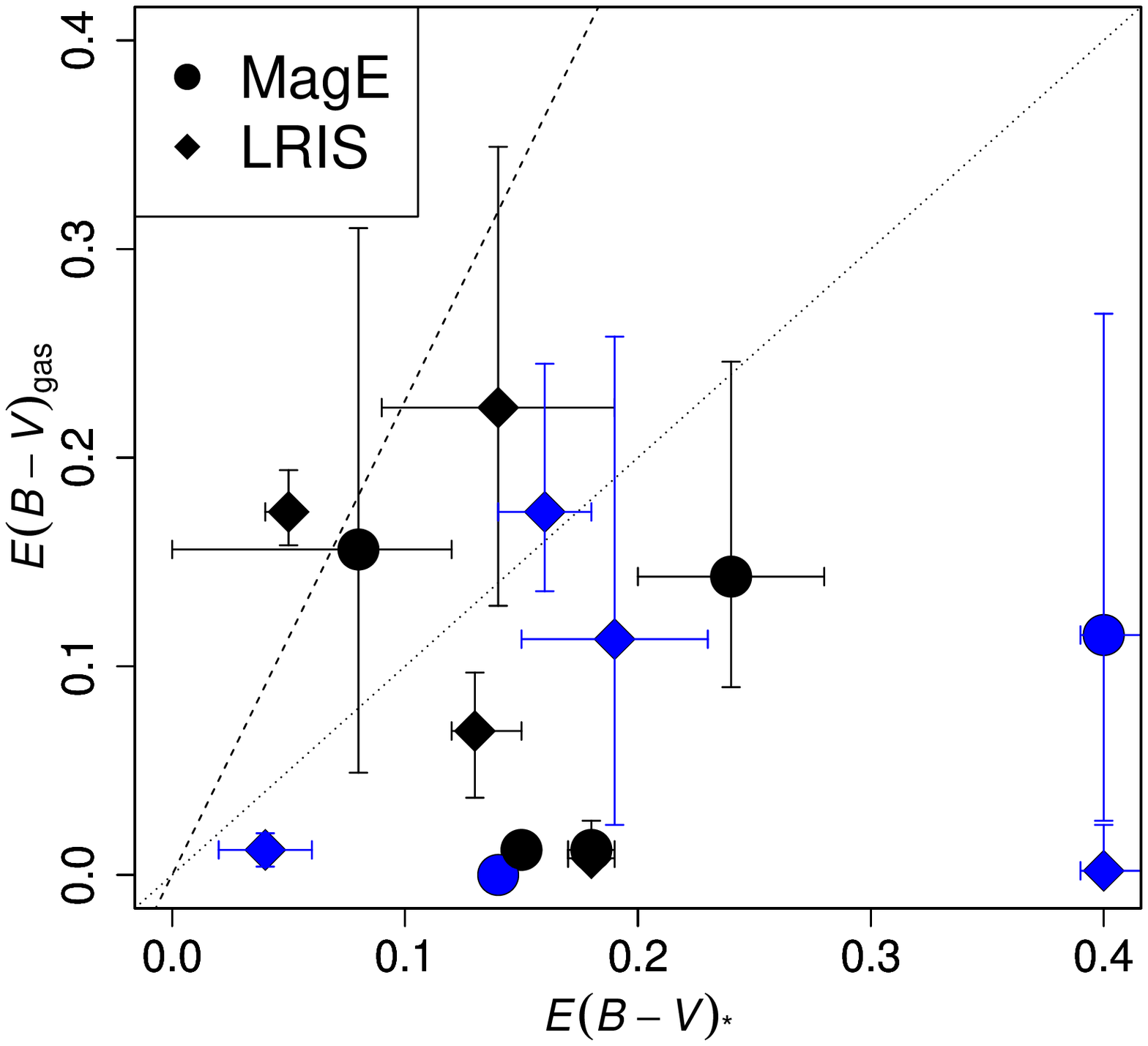}
\caption[]
{
$E(B-V)_{\rm gas}$ plotted against $E(B-V)_{\rm *}$ for all the sample. 
The circles show the MagE objects, while the diamonds are the LRIS objects. 
Blue color shows objects with a blue bump in the Ly$\alpha$ profile, 
while black color denotes those without. 
The dotted and dashed lines correspond to 
$E(B-V)_{\rm *}$ = $E(B-V)_{\rm gas}$ (\citealt{erb2006b}) 
and 
$E(B-V)_{\rm *}$ = $0.44 E(B-V)_{\rm gas}$ (\citealt{calzetti2000}), respectively.
}
\label{fig:comp_dust}
\end{figure}

The stellar dust extinction values, $E(B-V)_{*}$, for the sample 
have been derived in previous studies (\citealt{hashimoto2013, nakajima2013, shibuya2014b})
(see \S \ref{subsec:obs_results4-1}). 
Figure \ref{fig:comp_dust} compares them with gas dust extinction, $E(B-V)_{\rm gas}$, 
derived assuming the relation: 
\begin{eqnarray}
E(B-V)_{\rm gas}
\approx 
0.10 \tau_{\rm a}. 
\end{eqnarray}
Dotted and dashed lines correspond to empirical relations $E(B-V)_{*} = E(B-V)_{\rm gas}$ (\citealt{erb2006a}) 
and $E(B-V)_{*} = 0.44 E(B-V)_{\rm gas}$ (\citealt{calzetti2000}), 
respectively, for {\it host galaxies}. 
As \cite{kashino2013} have shown, the difference between 
$E(B-V)_{*}$ and $E(B-V)_{\rm gas}$ 
becomes smaller for higher-$z$ galaxies.

In this study, we expect that data points are located below these relations. 
This is because  $E(B-V)_{\rm gas}$ obtained from Ly$\alpha$ modeling 
is gas dust extinction for {\it outflowing shells}, 
which should be smaller than that for {\it host galaxies}. 
The figure shows that half of the sample roughly lie between the two lines,  
while the rest of the sample show low $E(B-V)_{\rm gas}$ values. 
A similar trend has been found in Figure 12 of \cite{verhamme2008} 
who have compared $E(B-V)_{*}$ and $E(B-V)_{\rm gas}$ for $z\sim3$ LBGs. 
They have assumed two different star formation histories (SFHs) 
in deriving $E(B-V)_{*}$:
a constant SFH indicated by red triangles
and an exponentially decreasing SFH indicated by blue open circles, 
the former of which is the same as that assumed in this study. 
Both our data and the red triangles in \cite{verhamme2006} 
are similarly distributed in the sense that half of the sample 
has comparable extinction values and the rest has low $E(B-V)_{\rm gas}$ values.

\subsubsection{FWHM(neb) vs. FWHM$_{\rm int}$(Ly$\alpha$)} \label{subsubsec:comp_fwhm}

\begin{figure}[]
\centering
 \includegraphics[width=8cm]{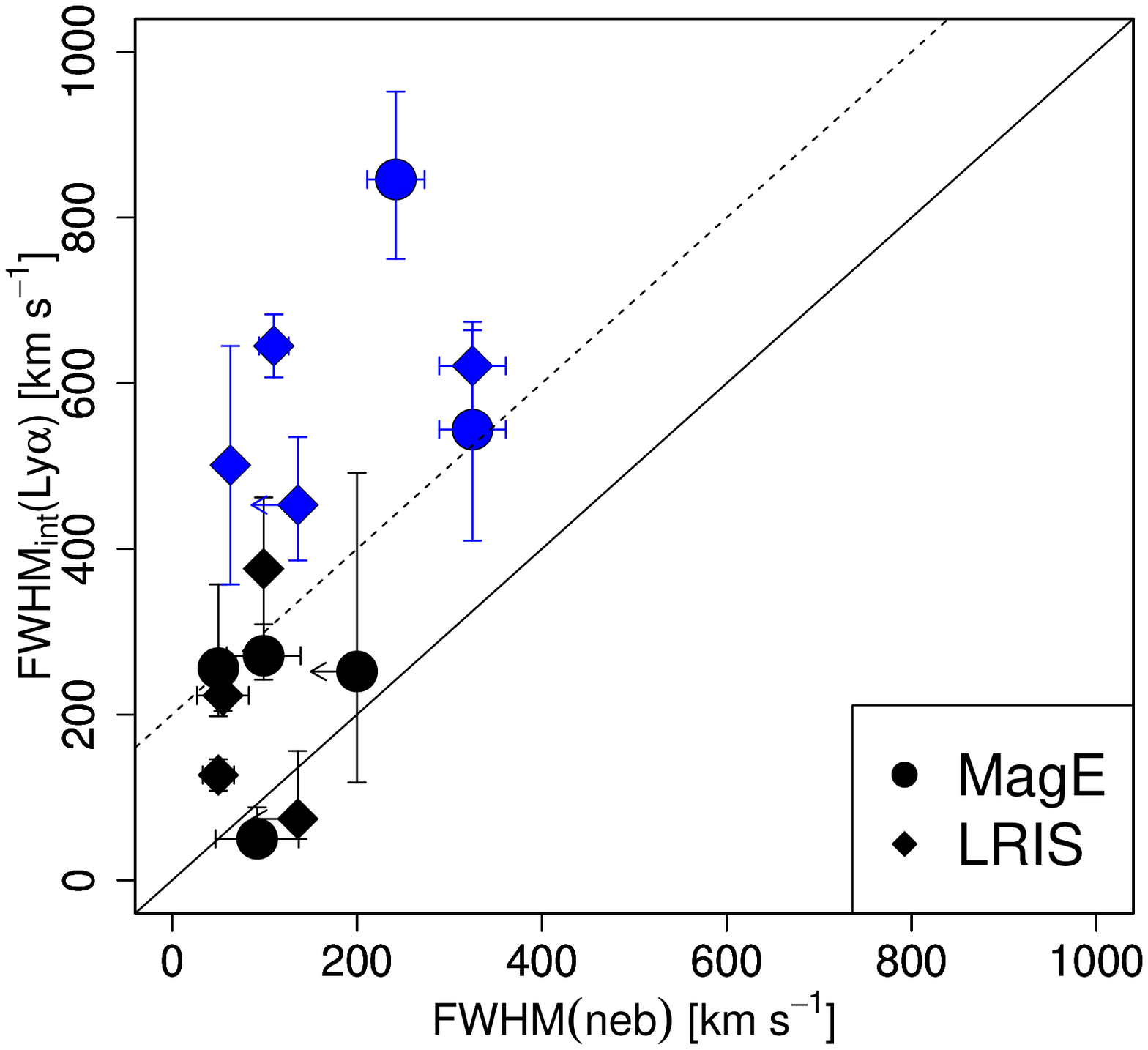}
\caption[]
{
FWHM$_{\rm int}$(Ly$\alpha$) plotted against FWHM(neb)
for all the sample. 
The meaning of the symbols and the colors is the same as in Figure \ref{fig:comp_dust}.
The solid line shows the one-to-one relation between the two FWHMs, 
while the dotted line is a relation between the two after 
taking into account an additional  scattering of Ly$\alpha$ photons 
by residual H$_{\rm I}$ atoms in the H$_{\rm II}$ region, 
FWHM$_{\rm int}$(Ly$\alpha$) = FWHM(neb) + 200. 
}
\label{fig:comp_fwhm}
\end{figure}

Figure \ref{fig:comp_fwhm} plots 
the observed FWHM of nebular emission lines, FWHM(neb), 
versus modeled FWHM of the intrinsic 
(i.e., before being affected by the radiative transfer effect) 
Ly$\alpha$ line, FWHM$_{\rm int}$(Ly$\alpha$). 
Assuming that both Ly$\alpha$ and nebular emission lines originate from {\sc Hii} regions, 
the two FWHMs should be similar. 
However, FWHM$_{\rm int}$(Ly$\alpha$) is systematically larger than  FWHM(neb). 
Additional scattering of Ly$\alpha$ photons in an H$_{\rm II}$ region 
due to residual H$_{\rm I}$ atoms in it may be at work. 
Assuming a static H$_{\rm II}$ region with a neutral hydrogen column density 
of log(N$_{\rm HI}$) $\lesssim17.0$ cm$^{-2}$, 
corresponding to an unity optical depth for ionizing photons, $\tau_{\rm ion} \lesssim 1$ 
(cf., \citealt{verhamme2015}), 
FWHM$_{\rm int}$(Ly$\alpha$) can be broadened by $200$ km s$^{-1}$ compared to FWHM(neb). 
As can be seen from Figure \ref{fig:comp_fwhm}, 
while this additional broadening would help explain the discrepancy 
for the non blue bump objects, 
it is still not enough for the blue bump objects. 
We discuss some interpretations for the huge FWHM$_{\rm int}$(Ly$\alpha$) 
in the blue bump objects in \S \ref{subsec:explanation_blue_bump}.

\begin{figure*}[]
\centering
 \includegraphics[width=18cm]{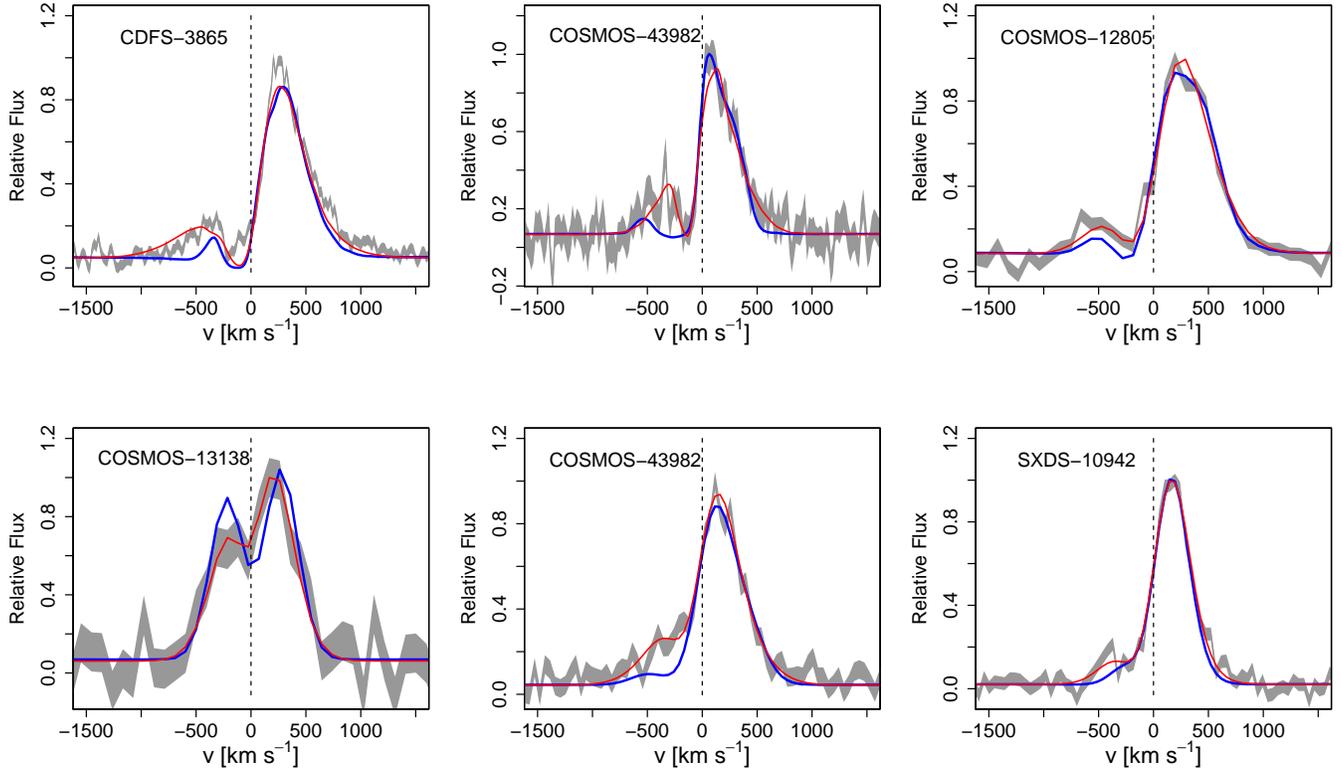} 
\caption[]
{
Ly$\alpha$ fittings with fixed FWHM$_{\rm int}$(Ly$\alpha$) = FWHM(neb) 
for the objects with a blue bump. 
The reproduced Ly$\alpha$ lines (blue) are overlaid on the observed ones (grey). 
For comparison, we also plot the reproduced profiles 
without fixing FWHM$_{\rm int}$(Ly$\alpha$) (red). 
}
\label{fig:fwhm-fixed_profile}
\end{figure*}

We also perform  Ly$\alpha$ profile fitting of the blue bump objects  
with fixing FWHM$_{\rm int}$(Ly$\alpha$) = FWHM(neb). 
As shown in Figure \ref{fig:fwhm-fixed_profile}, 
the blue bumps are poorly reproduced 
compared to the fitting without fixing FWHM$_{\rm int}$(Ly$\alpha$). 
We examine if  the derived best-fit model parameters differ between 
the free and fixed FWHM$_{\rm int}$(Ly$\alpha$) cases. 
While there is no systematic difference for $V_{\rm exp}$ and $N_{\rm HI}$, 
we find that $b$ ($\tau_{\rm a}$) becomes large (small) in the fixed FWHM$_{\rm int}$(Ly$\alpha$) case.
This would be related to the intrinsic degeneracy between them discussed in \S \ref{subsec:degeneracy}.

\subsubsection{EW(Ly$\alpha$) vs. EW$_{\rm int}$(Ly$\alpha$)} \label{subsubsec:comp_ew}

\begin{figure}[]
\centering
 \includegraphics[width=8cm]{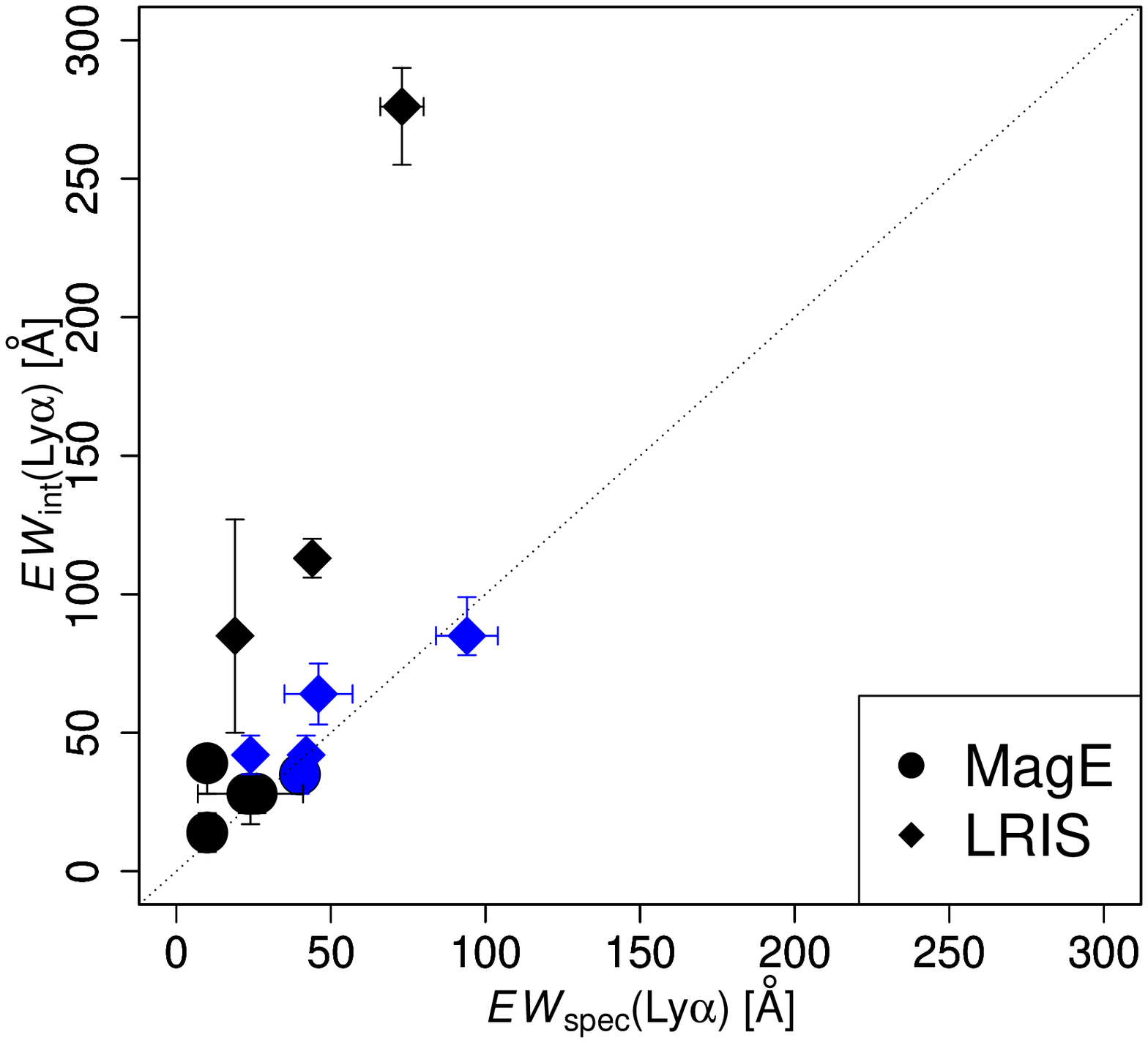}
\caption[]
{
EW$_{\rm int}$(Ly$\alpha$) plotted against EW(Ly$\alpha$) 
obtained from spectroscopy. 
The meaning of the symbols and colors is the same as in Figure \ref{fig:comp_dust}. 
Since we have assumed a uniform shell model which does not cause a EW(Ly$\alpha$) boost, 
the data points are expected to lie above the one-to-one relation. 
}
\label{fig:comp_ew_spec_model}
\end{figure}

Figure \ref{fig:comp_ew_spec_model} plots the observed EW(Ly$\alpha$) 
against the best fit intrinsic EW(Ly$\alpha$) obtained from the Ly$\alpha$ fitting, EW$_{\rm int}$(Ly$\alpha$).
Since we have modeled Ly$\alpha$ emission lines that fall in the slit, 
we use EW(Ly$\alpha$) values measured from spectra as the observed EW(Ly$\alpha$). 
All the data points are expected to lie above the one-to-one relation, 
EW$_{\rm int}$(Ly$\alpha$) $\gtrsim$ EW(Ly$\alpha$)$_{\rm spec}$. 
This is because we have used the uniform shell model 
which does not boost EW(Ly$\alpha$) unlike clumpy shell models 
(cf., \citealt{neufeld1991, laursen2013, duval2014, gronke2014}). 
As can be seen, all the data points satisfy the expectation within the 1$\sigma$ uncertainty.

\section{Discussion} \label{sec:discussion}

\subsection{Mystery of the Blue Bump Objects} \label{subsec:explanation_blue_bump}

As described in previous sections, 
FWHM$_{\rm int}$(Ly$\alpha$) $>$ FWHM(neb)
is required to well reproduce the Ly$\alpha$ profiles with the blue bump objects. 
As seen in Figure \ref{fig:fwhm-fixed_profile}, 
the position and flux of the blue bump 
are poorly reproduced if we fix FWHM$_{\rm int}$(Ly$\alpha$) =  FWHM(neb).

In this section, 
we first examine if there are any characteristic properties 
for the blue-bump objects, 
and discuss the origin of the large discrepancy 
between the two FWHMs.

\subsubsection{Any Difference in Properties between the Blue Bump and the Non Blue Bump Objects ? } \label{subsec:explanation_discrepancy_1}

In \S \ref{subsubsec:4-3-2}, 
we have argued that, among the model parameters, 
only FWHM$_{\rm int}$(Ly$\alpha$) is significantly different 
between the blue-bump objects the and non blue-bump objects. 
Here we examine the difference in 
stellar mass, Ly$\alpha$ luminosity, morphological ellipticity, 
and the merger fraction between the two samples.

First, it is possible that 
the non blue bump objects have faint Ly$\alpha$ luminosities and/or small stellar masses
so that the blue bump can not be observed. 
The Ly$\alpha$ luminosity of the blue bump sample 
ranges from L(Ly$\alpha$) = $0.3$ to $29.8$ $\times$ 10$^{42}$ erg s$^{-1}$ 
with a mean value of $8.8\pm5.6$ $\times$ 10$^{42}$ erg s$^{-1}$, 
whereas 
that of the non-blue bump sample ranges 
from L(Ly$\alpha$) = $0.5$ to $15.4$ $\times$ 10$^{42}$ erg s$^{-1}$ 
with a mean value of $7.0\pm2.1$ $\times$ 10$^{42}$ erg s$^{-1}$. 
This indicates that 
the two subsamples have similar Ly$\alpha$ luminosities. 
Likewise, the stellar mass of the  blue bump sample 
ranges  from log($M_{\rm *}$/$M_{\odot}$) = $7.73$ to $10.80$, 
with a mean value of $9.4\pm0.5$, 
whereas 
that of the non-blue bump sample 
ranges from log($M_{\rm *}$/$M_{\odot}$) = $7.84$ to $10.06$, 
with a mean value of $9.3\pm0.3$. 
Thus, this possibility is unlikely. 

Second, objects with a blue bump may be more likely to be seen edge-on than those without. 
Recent theoretical studies (\citealt{verhamme2012,zheng2014}) 
have investigated the inclination effects  
to the Ly$\alpha$ emissivity and profile. 
These studies have shown that the blue bump flux 
relative to the total Ly$\alpha$ flux is enhanced with an increasing ellipticity. 
Indeed, Ly$\alpha$ profiles seen edge-on  in these simulations 
resemble those produced by the static case of the spherical shell model. 
This is because outflowing gas is more likely to be blown out 
perpendicular to the galaxy disk, 
reducing the relative outflow velocity in the plane of the disk. 
As seen in Table \ref{tab:summary_other_prop}, 
there are three objects whose ellipticity has been measured. 
Due to the small number of objects, 
we cannot determine if there is any difference between 
the two subsamples.

Finally, as discussed in \cite{kulas2012} and \cite{chonis2013}, 
galaxy merging can be the origin of the blue bump. 
In this case, the redder and bluer Ly$\alpha$ emission components 
correspond to the two objects, respectively 
(see also  \citealt{cooke2010, rauch2011}). 
However, as described in \S \ref{subsec:obs_results4-3} (Table 4), 
the merger fraction in our sample is quite low.

We note here the observational results of \cite{erb2010} and \cite{heckman2011}. 
These studies have found that objects with a blue bump 
tend to have a low covering fraction of the neutral gas 
measured by LIS absorption lines. 
Unfortunately, we cannot test this trend with our sample 
because of a too small number of objects with detection of LIS absorption lines.

We conclude that there is no significant difference 
in Ly$\alpha$ luminosity, stellar mass, morphological ellipticity, 
or the merger fraction between the two samples. 
A large sample, whose Ly$\alpha$ and absorption line velocity properties 
as well as morphological and stellar population properties  
are simultaneously available, 
is needed to understand the origin of blue bumps.

\subsubsection{A Possible Explanation for Large FWHM$_{\rm int}$(Ly$\alpha$) in Blue Bunp Objects} \label{subsec:explanation_discrepancy_2}

In this subsection, we explore a possible explanation of the large discrepancy
between FWHM$_{\rm int}$(Ly$\alpha$) and FWHM(neb) in the blue bump objects.

It is possible that 
observed Ly$\alpha$ photons are produced 
not only from  
recombination of hydrogen gas ionized in H{\sc ii}  regions, 
but also from e.g., shock heating (\citealt{oti-floranes2012}),
fluorescence (e.g., \citealt{cantalupo2012, cantalupo2014}),  
and/or gravitational cooling (e.g., \citealt{dijkstra2006}). 
If these are taken into account, 
the huge FWHM$_{\rm int}$(Ly$\alpha$) in the blue bump objects 
could be explained as follows.

Fluorescence caused by a QSO would ionize the 
outer layer of the ISM of galaxies, 
and produce a large FWHM$_{\rm int}$(Ly$\alpha$). 
However, there are no QSOs around any of our objects.

Gravitational cooling is another mechanism that produces Ly$\alpha$ photons. 
When gas inflows into the gravitational potential well of a galaxy, 
the gravitational binding energy is converted into the thermal energy, 
which is in turn released as Ly$\alpha$ photons (e.g., \citealt{dijkstra2014}). 
Since it occurs in both the inner and outer regions of the galaxy, 
gravitational cooling can give a large FWHM$_{\rm int}$(Ly$\alpha$). 
Furthermore, gravitational cooling can reproduce 
not only the observed enhanced Ly$\alpha$ blue bump flux (e.g., \citealt{dijkstra2006}), 
but also the spatially extended Ly$\alpha$ source (\citealt{rosdahl2012}), 
i.e., diffuse Ly$\alpha$ haloes 
which are common features around galaxies 
\citep{steidel2011, matsuda2012, hayes2013, momose2014}. 
We note here that there exists a large uncertainty in modeling the Ly$\alpha$ 
emission from gravitational cooling due to its difficulty 
and assumed observation sensitivity 
(cf., \citealt{faucher-giguere2010, goerdt2010, rosdahl2012, yajima2012, yajima2015}). 
Our results as well as observational results quoted above can be useful for future modeling.

We conclude that 
not only the large discrepancy between the observed FWHM(neb) and FWHM$_{\rm int}$(Ly$\alpha$), 
but also the presence of a blue bump can be simultaneously explained 
if we introduce additional Ly$\alpha$ photons 
produced by gravitational cooling.

\subsection{Origin of Small $\Delta v_{\rm Ly\alpha ,r}$ in LAEs} 
\label{subsec:origin_small_vpeak}

As described in \S \ref{subsec:obs_results3}, 
the mean $\Delta v_{\rm Ly\alpha ,r}$ of LAEs, $\simeq 200$ km s$^{-1}$, 
is significantly smaller than that of LBGs, $\Delta v_{\rm Ly\alpha ,r} \simeq 400$ km s$^{-1}$
(LBGs: e.g., \citealt{steidel2010, rakic2011, kulas2012, schenker2013}, 
LAEs: e.g., \citealt{mclinden2011, hashimoto2013, chonis2013, shibuya2014b, erb2014}). 
We have also demonstrated that some LAEs have an extremely small $\Delta v_{\rm Ly\alpha ,r}$ 
of $\lesssim$ 100 km s$^{-1}$. 
\cite{hashimoto2013} and \cite{shibuya2014b} have also found that 
$\Delta v_{\rm Ly\alpha ,r}$ correlates with 
SFR, velocity dispersion, stellar mass, specific SFR, and dust extinction.
\cite{erb2014} have also found that 
$\Delta v_{\rm Ly\alpha ,r}$ correlates with velocity dispersion. 
In addition, they find that objects 
with a small $\Delta v_{\rm Ly\alpha ,r}$ 
have a large fraction of emission blueward of the systemic velocity, 
while the red wing of the Ly$\alpha$ profile and the outflow velocity 
traced by absorption lines remain unchanged. 
Following these findings, \cite{erb2014} have argued that 
the small $\Delta v_{\rm Ly\alpha ,r}$ in LAEs 
is consistent with a scenario 
where the opacity to Ly$\alpha$ photons is reduced 
by a bulk motion and/or covering fraction of the 
gas near the systemic velocity (see also \citealt{steidel2010}). 
These results suggest that 
$\Delta v_{\rm Ly\alpha ,r}$ is closely related 
with the physical size of the galaxy system. 
However, there are still no definitive conclusions why 
LAEs have small $\Delta v_{\rm Ly\alpha ,r}$.

In this subsection, 
we explore the origin of the small $\Delta v_{\rm Ly\alpha ,r}$ in LAEs 
using the largest sample of LAEs 
whose high-quality spectroscopy data and several properties have been obtained. 
There are several hypotheses 
which give $\Delta v_{\rm Ly\alpha, r}$ as small as $0-200$ km s$^{-1}$: 
a uniform shell ISM with a high-speed galactic outflow  (e.g., \citealt{verhamme2006}), 
a uniform shell ISM with a low neutral hydrogen column density 
(e.g., \citealt{verhamme2006, verhamme2015}), 
and other models such as a clumpy ISM 
with a low covering factor $f_{\rm c}$ (\citealt{hansen_oh2006, dijkstra2012, laursen2013}), 
or 
shell models with holes/cavities, i.e., $CF < 1$  (e.g., \citealt{behrens2014a, verhamme2015}). 
We quantitatively discuss the first two hypotheses based on 
our detailed comparison of data with uniform shell models 
(\S \ref{subsubsec:origin_small_vpeak_outflow} and \S \ref{subsubsec:low_NH}), 
then qualitatively discuss other models (\S \ref{subsubsec:origin_small_vpeak_other_models}).

\subsubsection{High Outflow Velocity} \label{subsubsec:origin_small_vpeak_outflow}

An outflow velocity larger than $V_{\rm exp} \sim 300$ km s$^{-1}$ 
can reduce $\Delta v_{\rm Ly\alpha ,r}$ because 
Ly$\alpha$ photons would drop out of resonance with {\sc Hi} atoms in the outflowing gas 
(e.g., \citealt{verhamme2006, verhamme2015}). 
However, 
our results of Ly$\alpha$ radiative transfer fitting in \S \ref{subsubsec:4-3-2} show 
that all objects have small $V_{\rm exp}$ of  $100 - 200$ km s$^{-1}$. 
Combined with the findings in \S \ref{subsubsec:comp_vexp} that 
these $V_{\rm exp}$ are consistent with the observed outflow velocities, $\Delta v_{\rm abs}$, 
we conclude that the high outflow velocity hypothesis is unlikely.

\subsubsection{Low $N_{\rm HI}$} \label{subsubsec:low_NH}

\begin{figure*}[]
\centering
 \includegraphics[width=15cm]{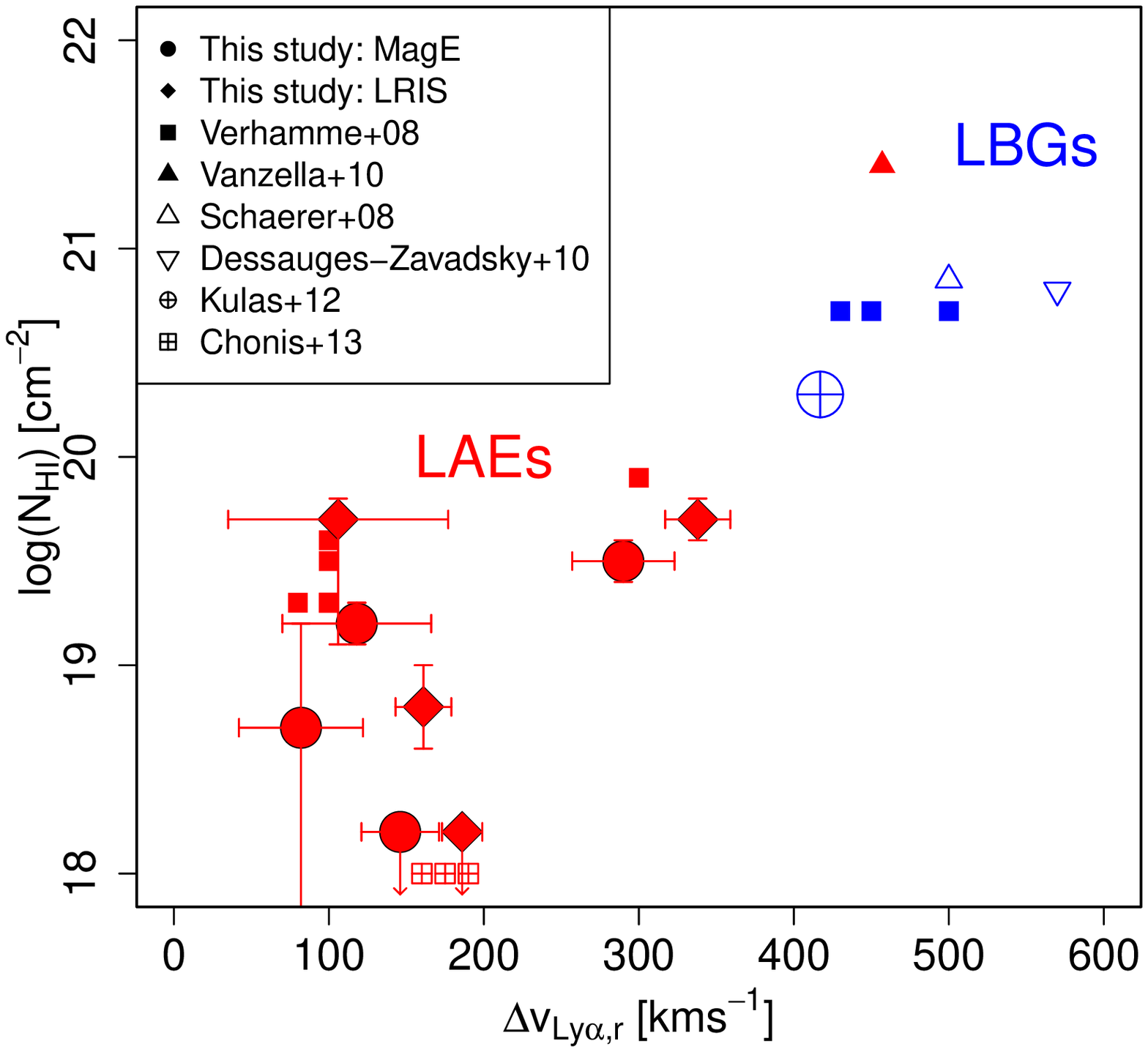}
\caption[]
{
log($N_{\rm HI}$) plotted against $\Delta v_{\rm Ly\alpha,r}$. 
Red (blue) symbols correspond to objects 
with EW(Ly$\alpha$)$_{\rm photo}$ larger (smaller) than $30$ \AA. 
Filled circles and diamonds are the non blue bump LAEs obtained by MagE and LIRS, respectively. 
Filled squares show  $z\sim3$ objects given by \cite{verhamme2008}. 
A filled triangle denotes a $z=5.56$ [{\sc Niv} emitter with EW(Ly$\alpha$) = 89\AA (\citealt{vanzella2010}). 
An open triangle is a lensed LBG with Ly$\alpha$ absorption, cB58 (\citealt{schaerer2008}), 
while an inverted triangle is a lensed LBG with Ly$\alpha$ absorption, 
``the 8 o'clock arc'' (\citealt{dessauges-zavadsky2010}).
In addition, the objects studied in \cite{kulas2012} and \cite{chonis2013} 
are plotted as a circle and three squares with a cross inside, respectively. 
For the purpose of display, 
three LAEs given by \cite{chonis2013} 
which have similar $\Delta v_{\rm Ly\alpha,r}$ and log($N_{\rm HI}$) values 
of 175 km s$^{-1}$ and 18 cm$^{-2}$, 
are shifted toward x-axis. 
We stress that all the data points and error bars are obtained 
assuming uniform expanding shells.  
The positions and/or error bars of data points in the figure can be significantly changed 
if we consider other models. 
See the text (\S \ref{subsubsec:low_NH}) for the detail. 
}
\label{fig:vpeak_nh}
\end{figure*}

We examine the low  $N_{\rm HI}$ hypothesis. 
Although it is difficult to directly measure $N_{\rm HI}$ in LAEs from observations, 
we have inferred it using the expanding shell model (\S \ref{subsubsec:4-3-2}). 
If we exclude the blue-bump objects from the sample, 
modeled Ly$\alpha$ profiles and parameters 
are all consistent with the  observed Ly$\alpha$ profiles and several fundamental observables. 
Thus, we consider the derived neutral hydrogen column density, $N_{\rm HI}$, 
to be reliable as well. 
Figure \ref{fig:vpeak_nh} is a plot of $\Delta v_{\rm Ly\alpha, r}$ against log($N_{\rm HI}$)
for the non-blue bump objects. 
We add results from the literature: 
\cite{verhamme2008}, \cite{schaerer2008}, \cite{vanzella2010}, and \cite{dessauges-zavadsky2010}. 
These authors have also utilized the model used in this study 
for $z\sim3$ LBGs with various EW(Ly$\alpha$) (\citealt{verhamme2008}), 
a strongly lensed LBG with Ly$\alpha$ absorption at $z\sim2.73$ (MS 1512-cB58) (\citealt{schaerer2008}), 
a peculiar $z=5.56$ [{\sc Niv} emitter with EW(Ly$\alpha$) = 89\AA (\citealt{vanzella2010}), 
and a lensed LBG with Ly$\alpha$ absorption, ``the 8 o'clock arc'' (\citealt{dessauges-zavadsky2010}).
We also add the results of \cite{kulas2012} and \cite{chonis2013}, 
although models used in these studies are different from the one used in this study. 
In the figure, 
objects with EW(Ly$\alpha$) $\gtrsim 30$ \AA\ are colored in red and labeled as LAEs, 
while those with EW(Ly$\alpha$) $< 30$ \AA\ are colored in blue and labeled as LBGs.

We caution readers that all the data points and error bars in Figure \ref{fig:vpeak_nh} 
are obtained assuming uniform expanding shell models 
(see \S \ref{subsec:fitting} for how we have obtained the error bars of our data points). 
The results can be significantly changed 
once if we consider other theoretical models such as clumpy or patchy models 
(see \S \ref{subsubsec:origin_small_vpeak_other_models}).

The figure shows a clear correlation between log($N_{\rm HI}$) and $\Delta v_{\rm Ly\alpha, r}$. 
As described in \S \ref{subsubsec:4-3-2}, 
the mean log($N_{\rm HI}$) in $z\sim2$ LAEs is log($N_{\rm HI}$) = 18.9 cm$^{-2}$,  
which is more than one order of magnitude lower 
than those of $z\gtrsim3$ LBGs, $\sim20.0$. 
We have excluded the 5 blue bump objects in our 12 LAEs 
in Figure \ref{fig:vpeak_nh} for a secure discussion. 
However, we note that they have comparable $N_{\rm HI}$ values 
to the non-blue bump objects, and are consistent with the correlation. 
We conclude that 
the small $\Delta v_{\rm Ly\alpha, r}$ in the LAEs 
can be well explained by the low $N_{\rm HI}$ hypothesis.

\subsubsection{Other Models} \label{subsubsec:origin_small_vpeak_other_models}

Throughout the paper, we have assumed a uniform expanding shell models 
constructed by \cite{verhamme2006} and \cite{schaerer2011}. 
In this section, we qualitatively discuss alternative models such as clumpy and patchy models. 
\cite{hansen_oh2006} have analytically investigated the Ly$\alpha$ radiative  transfer in multi-phase media, 
especially, the one through dusty optically thick gas clumps. 
They have shown that radiative transfer strongly depends on covering factor, $f_{\rm c}$  
(see also \citealt{dijkstra2012, laursen2013}). 
As can be seen from Figure 20 of \cite{hansen_oh2006}, 
asymmetric Ly$\alpha$ profiles with small $\Delta v_{\rm Ly\alpha,r}$ can be reproduced 
by clumpy models with low $f_{\rm c}$, i.e., 
a small average number of interaction for Ly$\alpha$ photons before escaping from the galaxy. 
Recently, 
\cite{behrens2014a} and \cite{verhamme2015} have investigated 
the Ly$\alpha$ radiative transfer in a shell with holes and/or cavities, 
i.e., $CF < 1$. 
These studies have shown that, if a shell has holes,  
the modeled Ly$\alpha$ line profile through a transparent line of sight has $\Delta v_{\rm Ly\alpha ,r} = 0$ km s$^{-1}$ 
even if convolved with spectral resolutions used for the observations 
(see Figures 4 and 5 in \citealt{verhamme2015}).
We qualitatively test the hypothesis by comparing Ly$\alpha$ and nebular emission line profiles.  
In the case of a patchy ISM, if we observe the galaxy through a transparent line of sight,
observed Ly$\alpha$ line profiles would be indistinguishable 
from nebular emission line profiles. 
This is because the main Ly$\alpha$ component is dominant and is not affected 
by the radiative transfer effect. 
As can be seen in Figure 1, 
COSMOS-08357 and COSMOS-43982 have 
indistinguishable Ly$\alpha$ and nebular line profiles. 
We conclude that 
at least two objects, COSMOS-08357 and COSMOS-43982, 
could be explained by patchy ISM models.

Thus, we have demonstrated that 
the small $\Delta v_{\rm Ly\alpha,r}$ in LAEs can be also well reproduced 
by clumpy and/or patchy ISM. 
However, future detailed Ly$\alpha$ modeling 
assuming clumpy and patchy ISM 
are needed for a more definitive conclusion. 
Whichever hypothesis is the most relevant one, 
the key for the small $\Delta v_{\rm Ly\alpha ,r}$ in LAEs 
would be the reduced number of resonant scattering of Ly$\alpha$ photons.

Hereafter in this section, 
we focus on the results obtained from uniform expanding shell models with low $N_{\rm HI}$.

\subsection{Interpretation of Low $N_{\rm HI}$ in LAEs} \label{subsec:interpretation_impplication_low_NH}

We have shown that, on the assumption of uniform shell models, 
the most likely situation for the smaller $\Delta v_{\rm Ly\alpha,r}$ 
in the present sample is that LAEs have low $N_{\rm HI}$. 
We can envisage three possible scenarios for LAEs having low $N_{\rm HI}$.

First, it is possible that LAEs have a low {\sc Hi} gas mass. 
Indeed, \cite{pardy2014} have detected an {\sc Hi} 21cm line 
for 14 local galaxies with Ly$\alpha$ emission 
(Lyman Alpha Reference Sample; \citealt{ostlin2014}), 
to find that the derived {\sc Hi} gas mass tentatively anti-correlates with EW(Ly$\alpha$). 
This trend is also consistent with theoretical predictions 
(private communications with T. Garel and C. Lagos).

Second, if a galaxy has a high gas ionization state, 
ionizing photons would efficiently ionize the neutral gas in the ISM. 
This would decrease the thickness of the {\sc Hi} gas in the outflowing shell, 
and lower their $N_{\rm HI}$. 
This picture is consistent with the recent finding of \cite{nakajima2014} 
that LAEs have a significantly higher ionization state than that of LBGs 
at the same redshift. 
The high ionization state in LAEs would be due to their young stellar populations 
(e.g., \citealt{pirzkal2007, ono2010a, ono2010b}). 
Young O- and B type stars in LAEs would efficiently produce ionizing photons, 
and reduce the $N_{\rm HI}$ of the surrounding gas.

Finally, in the case of a face-on galaxy, 
we would see a lower $N_{\rm HI}$ 
because Ly$\alpha$ photons would experience 
a shorter path length out of the disk  (e.g., \citealt{verhamme2012, zheng2014}).
Indeed, \cite{shibuya2014a} have statistically examined 
the ellipticity, an indicator of the inclination, for $z\sim2$ LAEs using $HST$ data. 
A weak trend has been found that high EW(Ly$\alpha$) objects 
are less inclined.

A combination of these effects would reduce 
the number of resonant scatterings of Ly$\alpha$ photons. 
This would, in turn, decrease the Ly$\alpha$ velocity offset, $\Delta v_{\rm Ly\alpha ,r}$ 
(e.g., \citealt{mclinden2011, hashimoto2013, erb2014}) 
and the Ly$\alpha$ spatial offset, $\delta_{\rm Ly\alpha}$
(\citealt{jiang2013, shibuya2014a}).

\subsection{Implication of Small $\Delta v_{\rm Ly\alpha ,r}$ and $\Delta v_{\rm peak}$ in LAEs} \label{subsec:implication_small_vpeak}

Recent theoretical studies (\citealt{behrens2014a, verhamme2015}) have proposed 
that the Ly$\alpha$ line profile can be used as a probe of 
Lyman continuum (LyC; $\lambda < 912$\AA) leaking galaxies (LyC leakers). 
LyC leakers are thought to have contributed 
to cosmic reionization. 
Observationally, detections of LyC emission are claimed for LAEs and LBGs  
both spectroscopically (e.g., \citealt{shapley2006})
and photometrically (e.g., \citealt{iwata2009} and \citealt{nestor2013}). 
However, the success rate is very low 
possibly because LyC leakers are extremely faint objects (e.g., \citealt{ouchi2008}).

\cite{verhamme2015} have investigated two scenarios for the ionizing photon escape: 
(1) the density bounded {\sc Hii} regions with an extremely low $N_{\rm HI}$ value, log($N_{\rm HI}$) $\lesssim$ 17.0 cm$^{-2}$, 
corresponding to an unity optical depth for ionizing photons, 
(2) or a galaxy has a partial spatial covering fraction of the gas. 
They have shown that 
$\Delta v_{\rm Ly\alpha, r}$ ($\Delta v_{\rm peak}$, if the blue bump exists) 
is extremely small in these cases, 
$\Delta v_{\rm Ly\alpha, r} \lesssim 100$ km s$^{-1}$ 
($\Delta v_{\rm peak} \lesssim 150$ km s$^{-1}$) 
(see also \citealt{jaskot2014,martin2015}).

On the other hand, as described in \S \ref{subsec:obs_results3}, 
objects with a large EW(Ly$\alpha$) value 
tend to have a  small $\Delta v_{\rm Ly\alpha, r}$ ($\Delta v_{\rm peak}$), 
implying that they are good candidates of LyC leakers. 
If the ionizing photon escape fraction, $f_{\rm esc}$,  is high, 
the recombination Ly$\alpha$ line might be weaken. 
However, as \cite{nakajima2014} have shown, 
EW(Ly$\alpha$) can be as high as 100 \AA\ even at $f_{\rm esc} \sim 0.5$. 

Thus, we propose that 
selecting objects with EW(Ly$\alpha$) as large as 100\AA\ is a promising way 
to search for LyC leaking galaxies. 
The merit of this candidate selection technique is that it does not require spectroscopy.

\section{SUMMARY AND CONCLUSION} \label{sec:conclusions}

We have presented the results of a Ly$\alpha$ profile analysis of 
twelve LAEs at $z\sim2.2$ 
for which high spectral resolution Ly$\alpha$ lines are obtained 
in \cite{hashimoto2013} and \cite{shibuya2014b} with 
Magellan/MagE or Keck/LRIS. 
Two objects have been observed with both spectrographs.
All twelve objects have detections of nebula emission lines 
which are used not only to define the systemic redshift 
but also to infer the intrinsic FWHM of the Ly$\alpha$ line. 
We have also derived the galactic outflow velocity 
from blueshifted low-ionization state (LIS) metal absorption lines 
with respect to the systemic redshift 
for three individual LRIS spectra 
as well as for a stacked spectrum of four MagE spectra. 
In addition, we have obtained stellar dust extinction from SED fit. 
The high spectral resolution Ly$\alpha$ data 
in conjunction with these measurements 
have enabled us to perform detailed comparisons 
between observed and modeled Ly$\alpha$ lines.   
Our main results are as follows.

\begin{itemize}
\item 
We find that all 12 objects have Ly$\alpha$ profiles 
with a main peak redward of the systemic redshift 
and five objects (six spectra) have a weak, secondary peak 
blueward of the systemic redshift (the blue bump). 
For a sample of 17 objects from our study and the literature 
with a resolved Ly$\alpha$ line,  
we estimate the ratio of LAEs with a blue bump to be $\sim50\%$, 
which is slightly higher than that of LBGs. 
We have obtained 
$\Delta v_{\rm Ly\alpha ,r} = 174\pm19$ km s$^{-1}$ 
($\Delta v_{\rm Ly\alpha ,b} = -316\pm45$ km s$^{-1}$), 
which is smaller than (comparable to) that of LBGs, 
$\Delta v_{\rm Ly\alpha ,r} \simeq 400$ km s$^{-1}$ 
($\Delta v_{\rm Ly\alpha ,b} = -367\pm46$  km s$^{-1}$).

\item 
The high spectral resolution and sensitivity of Subaru/FMOS 
have enabled us to detect 
two-component [{\sc Oiii}] profiles in two LAEs for the first time.
While its origin is not clear, 
we find that even the FWHM of the broad component 
is as small as $70-80$ km s$^{-1}$. 
This excludes the possibility of its origin being AGN activity or powerful hot outflows.

\item 
We have applied the uniform expanding shell model 
constructed by \cite{verhamme2006} and \cite{schaerer2011} 
to our sample. 
The model successfully reproduces 
not only Ly$\alpha$ profiles but also 
the galactic outflow velocity measured from LIS absorption lines 
and the FWHM of nebular emission lines 
for the non blue-bump objects. 
However, for the blue-bump objects, 
the intrinsic FWHMs of Ly$\alpha$ predicted by the model 
is significantly larger than 
the observed FWHMs of nebular emission lines.

\item 
For the blue bump objects, we have tried another fit 
fixing the intrinsic FWHM of Ly$\alpha$ 
to the observed FWHM of nebular emission lines. 
The position and flux of the blue bump are poorly reproduced.

\item 
To understand the large discrepancy between 
FWHM$_{\rm int}$(Ly$\alpha$) and FWHM(neb) 
in the blue-bump objects, 
we have examined if objects with and without a blue bump 
have different properties 
such as the Ly$\alpha$ luminosity, stellar mass, 
and the merger fraction. 
We find no significant difference   
between the two samples. 
We propose that taking into account Ly$\alpha$ photons produced 
by gravitational cooling might simultaneously explain 
the large FWHM$_{\rm int}$(Ly$\alpha$) 
and the existence of a blue bump.

\item
We quantitatively demonstrate that  
the small $\Delta v_{\rm Ly\alpha ,r}$ in LAEs 
can be well explained by uniform expanding shell models 
with neutral hydrogen column density 
as low as log($N_{\rm HI}$) = 18.9 cm$^{-2}$. 
This value is 
more than one order of magnitude lower than that of LBGs, 
and is consistent with the recent findings that LAEs 
have a high ionization parameter  
and a low {\sc Hi} gas mass. 
These results imply that 
low $N_{\rm HI}$ is the key 
for the small $\Delta v_{\rm Ly\alpha ,r}$ 
as well as the Ly$\alpha$ escape mechanism. 
However, we caution readers that our results are based only on 
uniform expanding shell models, 
and that future detailed modeling with clumpy and/or patchy ISM 
are needed for a definitive conclusion.

\item 
As an implication of the small $\Delta v_{\rm Ly\alpha, r}$ 
and low $N_{\rm HI}$ in high EW(Ly$\alpha$) objects, 
we propose that targeting high EW(Ly$\alpha$) objects 
would be an efficient way to search for Lyman Continuum leaking galaxies 
from photometry data alone. 

\end{itemize}

\section*{Acknowledgements}

We thank an anonymous referee 
for valuable comments that have greatly improved the paper. 
We are grateful to 
Thibault Garel, Claudia Lagos,  Ivana Orlitov\'a, 
Kentaro Motohara, Nobunari Kashikawa, Ryohei Kawabe, Kohtaro Kohno, and Toru Yamada 
for their helpful comments and suggestions. 
We acknowledge Richard Ellis and Matthew Schenker for kindly providing us with their LBG data. 
We also thank George Becker for making the software package $\tt{MagE\_REDUCE}$ available to us. 
This work was supported by 
World Premier International Research Center Initiative
(WPI Initiative), MEXT, Japan, 
and KAKENHI (23244022) and (23244025) 
Grant-in-Aid for Scientific Research (A)
through Japan Society for the Promotion of Science (JSPS).
T.H. also acknowledges the JSPS Research Fellowship for Young Scientists. 
A.V. was supported by a Fellowship ``Boursi\`ere d'Excellence'' of Geneva University.
M.R. was supported by a grant AST-1108815 from the National Science Foundation.
%
\appendix

Figures 12 - 14 show 2D $\chi^{2}$ contours for CDFS-3865. 
All the grids are shown in grey dots. 
The blue (red) grids in these figures show 
those satisfying $\Delta \chi^{2} \leq 6.17 \ (11.8)$ 
above the raw minimum $\chi^{2}$ designated by the white dots, 
i.e., the 3 (5) $\sigma$ uncertainty in \ the parameters (\citealt{press1992}).

\begin{figure*}[]
\centering
 \includegraphics[width=11cm]{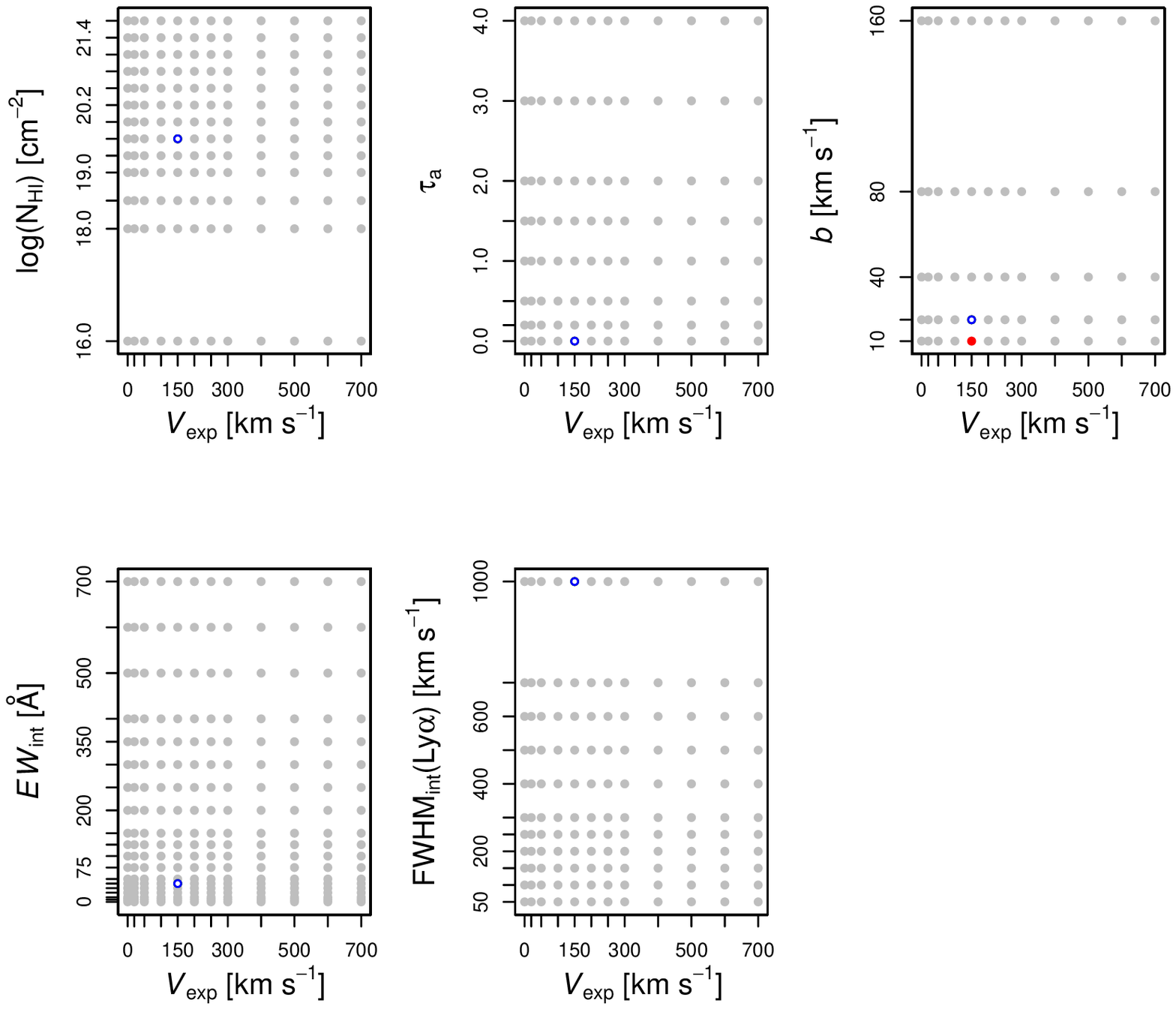}
 \includegraphics[width=11cm]{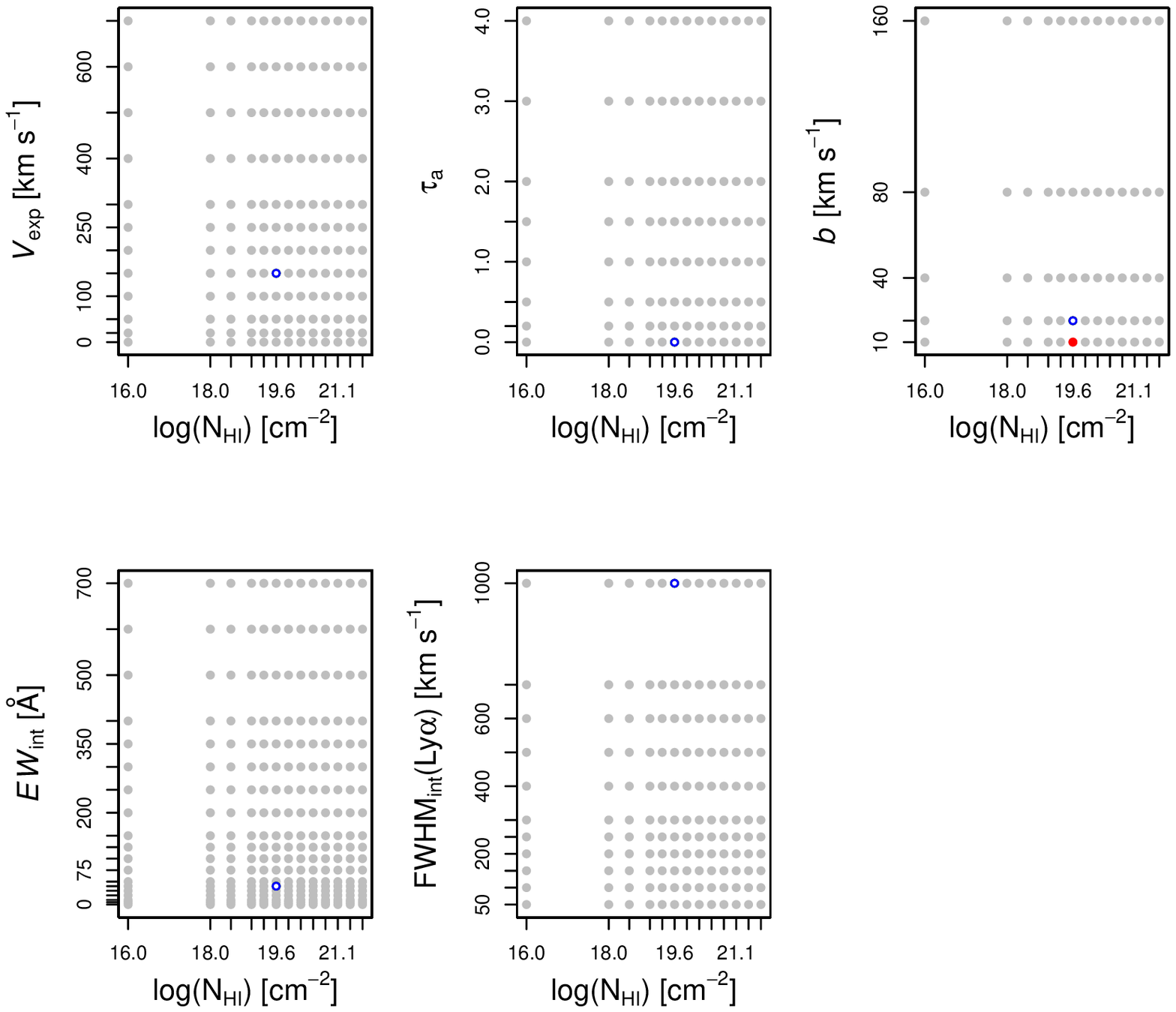}
\caption[]
{
Upper and lower five panels show 2D $\chi^{2}$ contours for $V_{\rm exp}$ 
and log($N_{\rm HI}$), respectively, for CDFS-3865. 
The grids colored with blue (red) denote 
those within the 3 (5) $\sigma$ level from 
the minimum $\chi^{2}$ grid shown as a white dot. 
}
\label{fig:two_component_RT}
\end{figure*}

\begin{figure*}[]
\centering
 \includegraphics[width=11cm]{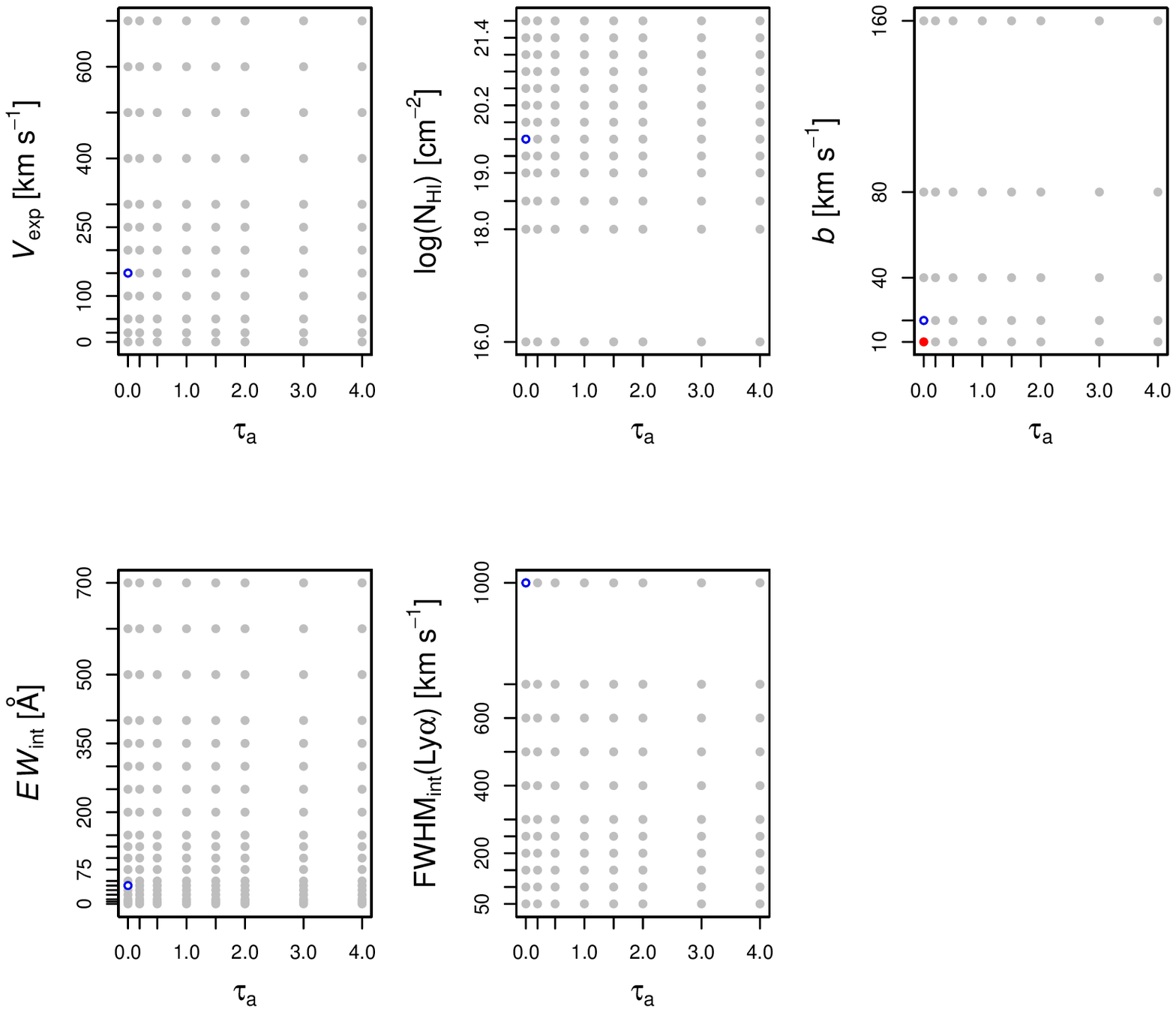}
 \includegraphics[width=11cm]{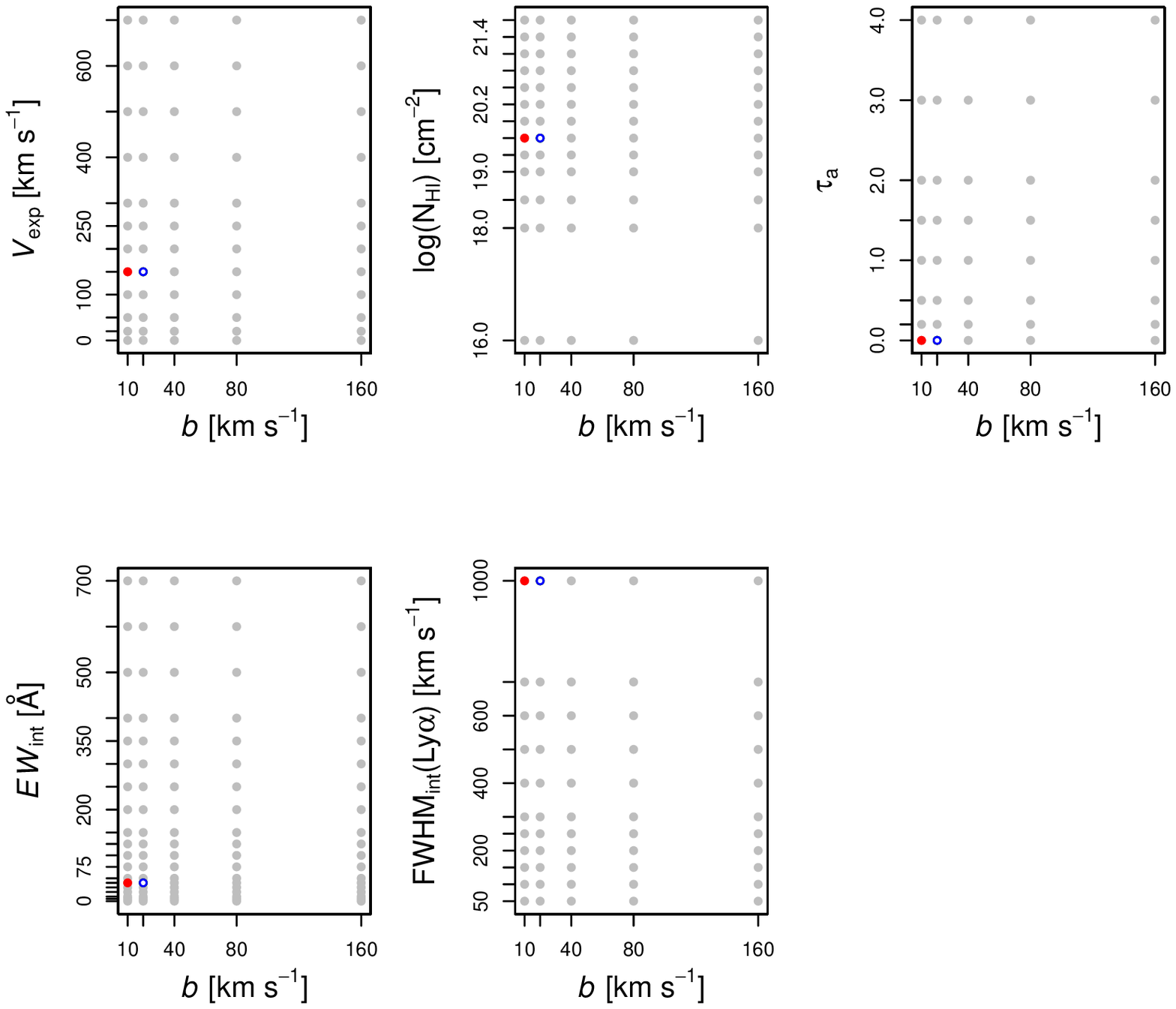}
\caption[]
{
Upper and lower five panels show 2D $\chi^{2}$ contours for $\tau_{\rm a}$ 
and $b$, respectively, for CDFS-3865. 
The meaning of the colors is the same as in Figure 12. 
}
\label{fig:two_component_RT}
\end{figure*}

\begin{figure*}[]
\centering
 \includegraphics[width=11cm]{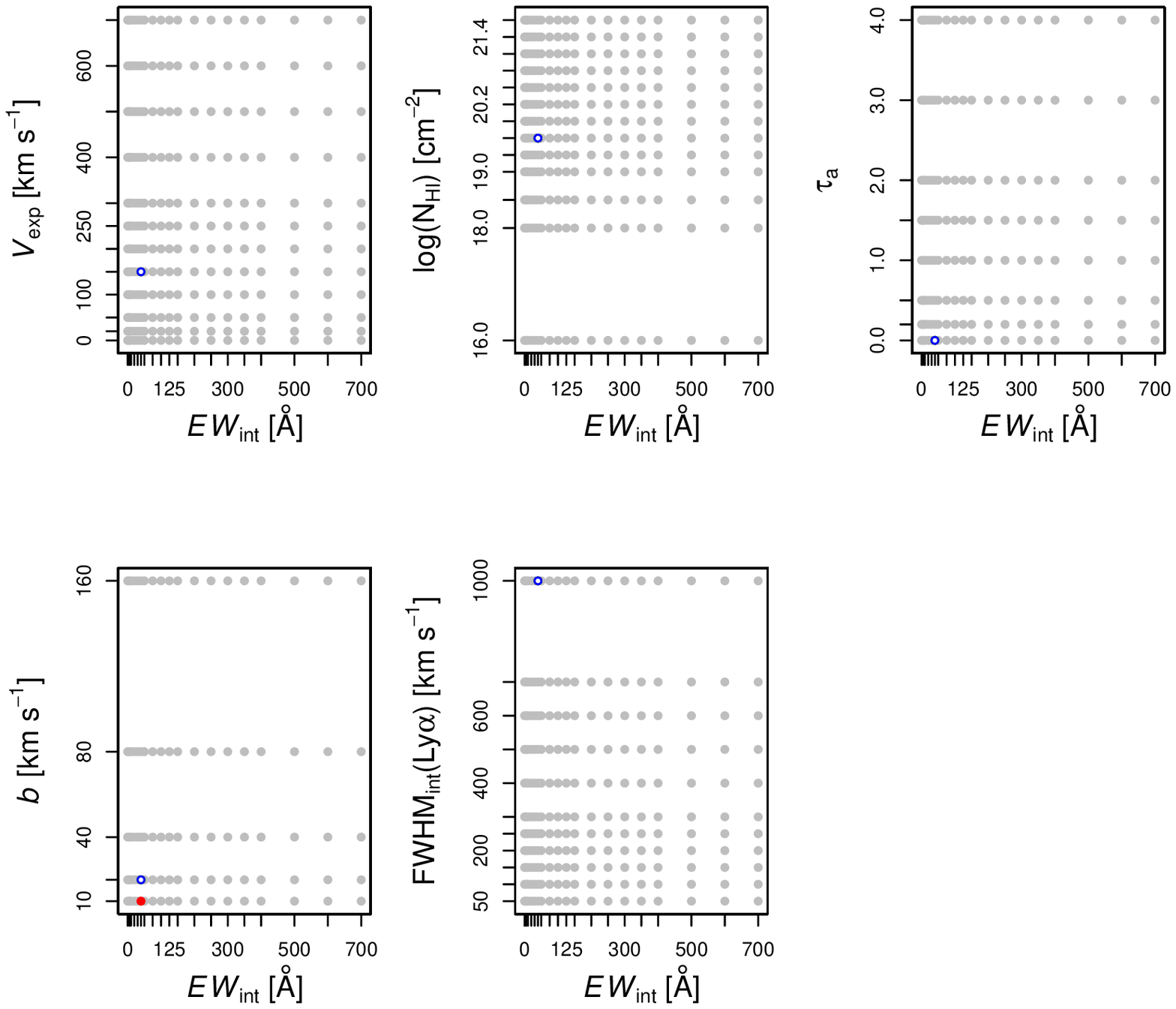} 
 \includegraphics[width=11cm]{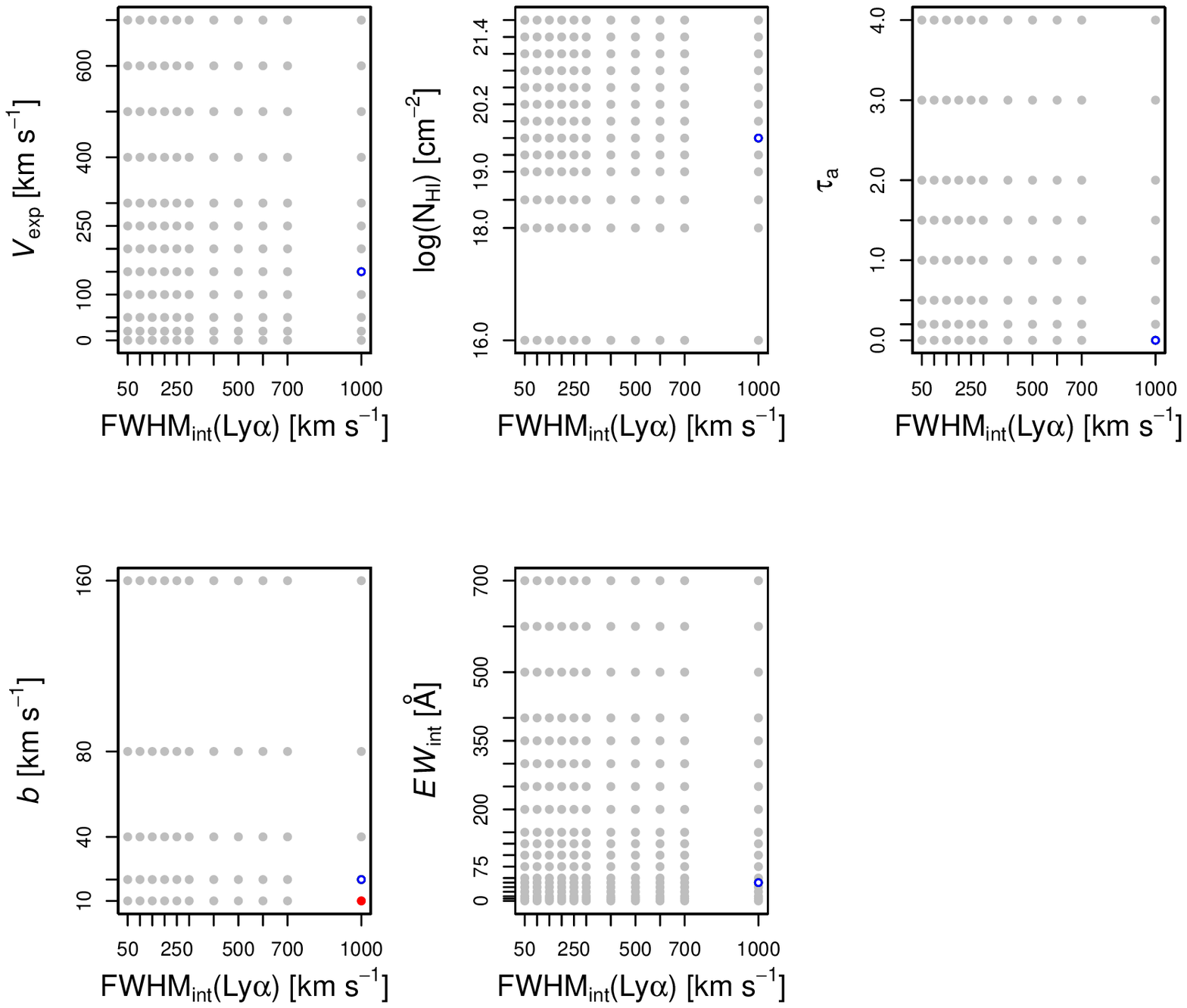}
\caption[]
{
Upper and lower five panels show 2D $\chi^{2}$ contours for EW$_{\rm int}$(Ly$\alpha$)
and FWHM$_{\rm int}$(Ly$\alpha$), respectively, for CDFS-3865. 
The meaning of the colors is the same as in Figure 12. 
}
\label{fig:two_component_RT}
\end{figure*}

\bibliographystyle{apj}
\bibliography{hashimoto2015}

\end{document}